\RequirePackage{fix-cm,amssymb}
\documentclass[smallextended]{svjour3}       
\smartqed  

\usepackage{graphicx}
\usepackage{url,color}
\usepackage{comment}
\usepackage{caption}
\usepackage{subcaption}
\begin{document}

\title{SARS-CoV-2 dissemination using a network of the United States counties \thanks{Ruriko Yoshida is supported partially from NSF Statistics Program DMS 1916037.}
}


\author{
        Patrick Urrutia \and
David Wren         \and
Chrysafis Vogiatzis \and
        Ruriko Yoshida 
}


\institute{
            P.\ Urrutia \at
            Department of Operations Research\\
              Naval Postgraduate School\\
              \email{patrick.urrutia@nps.edu}   
        \and
        D.\ Wren  \at
              Department of Operations Research\\
              Naval Postgraduate School\\
              \email{david.m.wren.mil@mail.mil}           
           \and
           C.\ Vogiatzis \at 
                Industrial and Enterprise Systems Engineering\\
                University of Illinois at Urbana-Champaign\\
                \email{chrys@illinois.edu}
           \and
           R.\ Yoshida \at
            Department of Operations Research\\
              Naval Postgraduate School\\
              \email{ryoshida@nps.edu}   
}

\date{Received: date / Accepted: date}

\maketitle

\begin{abstract}
During 2020 and 2021, severe acute respiratory syndrome coronavirus 2 (SARS-CoV-2) transmission has been increasing amongst the world's population at an alarming rate. Reducing the spread of SARS-CoV-2 and other diseases that are spread in similar manners is paramount for public health officials as they seek to effectively manage resources and potential population control measures such as social distancing and quarantines. By analyzing the United States' county network structure, one can model and interdict potential higher infection areas. County officials can provide targeted information, preparedness training, as well as increase testing in these areas. While these approaches may provide adequate countermeasures for localized areas, they are inadequate for the holistic United States. We solve this problem by collecting coronavirus disease 2019 (COVID-19) infections and deaths from the Center for Disease Control and Prevention{\color{black},} and {\color{black} adjacency between all counties obtained} from the United States Census Bureau. Generalized network autoregressive (GNAR) time series models have been proposed as an efficient learning algorithm for networked datasets. This work fuses network science and operations research techniques to univariately model COVID-19 cases, deaths, and current survivors across the United States' county network structure.
\end{abstract}

\section{Introduction}
\label{intro}

The severe acute respiratory syndrome coronavirus (SARS-CoV-2) that was first detected in late 2019 and the resultant disease (COVID-19) have completely upended the way we {\color{black}led} our lives. The global pandemic has lead to debilitating damage to our lives, economies, health care systems, and food security \cite{Impactof55:online}. The virus is extremely transmissible, spreading through water droplets produced when talking, coughing, or sneezing \cite{CDCspread}. Since the virus' first reports in 2019, it has mutated multiple times and some of these mutations have proven to spread faster and infect easier \cite{Yang2021.06.21.21259268}. Additionally, when a virus mutates, combating it with vaccines and other public health measures becomes increasingly difficult \cite{Sanjuan}. Hence, it is in our best interest to investigate the spreading mechanisms and obtain ways to predict outbreaks, which are one of the goals of this study. The transmission of the virus has been further aided by modern society's hypermobility \cite{musselwhite2020editorial}. Early on, COVID-19 spread fast within China and other countries due to both international and national travel, through air, land, and seas.  

{\color{black}In previous works, the relationship between human mobility and epidemics spreading has been investigated. Recently, we have had some very important works relating COVID-19 to traffic. Indicatively, we mention the fundamental work of Kraemer et al. \cite{doi:10.1126/science.abb4218} which analyzed human mobility data and traced infection metrics in the early and later stages of the COVID-19 pandemic. The results reinforce that earlier in the pandemic, strict travel restrictions are helpful and lead to easier confinement and control; later, once the outbreak is spread, travel restrictions are less useful and local measures, such as social distancing and masking, are preferable. In a second recent work \cite{wu2021traffic}, the authors posit that while traffic and human mobility are often the culprits for driving viral spreading, the traffic network structure is often overlooked from these studies. Hence, they propose a traffic-driven model that accounts for that; here, we also account for human mobility through the transportation network and consider the effects of traffic to the spread of a disease.}

{\color{black}Specifically, in this work we also investigate whether and how travel patterns affect COVID-19 dissemination; we do so by employing generalized network autoregression on a proxy of the United States transportation network.} We generate and use a network of all counties in the United States in an effort to forecast the spread of COVID-19 using data available for each county as well as travel patterns across counties. The remainder of the manuscript is organized as follows. First, in Section \ref{sec:lit_review}, we provide a brief literature review on models that have been put to the use to forecast COVID-19 cases and to protect the communities from its transmission. Then, in Section \ref{sec:methods}{\color{black},} we discuss our approach using the generalized network autoregression (GNAR). We also provide a description of the data that {\color{black}were} acquired to perform the analysis in Section \ref{sec:data}. Section \ref{sec:results} presents the computational experiments and the results we observed during our analysis. We conclude this work in Section \ref{sec:conclusion}.

\section{Literature review}
\label{sec:lit_review}

Due to the impact of COVID-19 in our daily lives, a lot of research has already appeared {\color{black}on} the analysis of the spread of the disease. That said, epidemics and pandemics such as the one caused by SARS-CoV-2 are not a recent phenomenon for humanity. As an example, the ``Spanish flu'' ravaged the world in the early 1900s. Shortly after the outbreak of the disease, in 1918, researchers  Kermack and  McKendrick published papers that presented mathematical models for predicting the number of infections in a population as a function of time: the assumption was that it is valid to split the population into smaller {\color{black} clusters or} ``compartments'' when analyzing a disease's propagation through a population \cite{Kermack:1991}. Their foundational work continues to help epidemiologists model outbreaks of diseases today.

More recently, epidemiological models such as the Susceptible-Infected-Removed (Recovered) (SIR) and Susceptible-Exposed-Infected-Removed (Recovered) (SEIR) and other extensions have been put to use to model the movement of individuals from one ``compartment'' {\color{black} (i.e., Susceptible, Exposed, Infected, Removed)} to the next \cite{rodrigues2016application}. {\color{black} As an example, a person may be moved from the initial state of Susceptible to the intermediate state of Infected upon exposure to and infection with a disease; later that same person may be categorized as Removed once they recover.} As expected, such epidemiological models have been applied in the fight against COVID-19. These models have been largely successful, revealing their utility for policy to prevent the spread of disease. 

In Cameroon, research based on SIR determined that the number of COVID-19 cases was limited due to the health precautions taken \cite{Nguemdjo:2020}. Another similar application of the SIR model originates from Saudi Arabia, where researchers analyzed the number of COVID-19 cases and deaths both with and without public health measures such as quarantine enforcement \cite{Singh:2020}. Although SIR models have been accurate enough in predicting the size of the COVID-19 outbreaks, more recent research indicates that individuals who contract the virus once can become infected again \cite{Gallagher:2021}, necessitating a means to dynamically update the parameters of the SIR model in an effort to improve its predictive power. In \cite{Chen}, the authors propose {\color{black}time-varying} these parameters to account for changes over time, using machine learning to determine exactly how to update these parameters. 

Moreover, the incubation period of COVID-19 (i.e., the period during which an infected individual bears no symptoms yet can still transmit the virus to others) has proven to be an important factor in the spread of COVID-19 \cite{Hoehl:2020}. While asymptomatic, some recently infected individuals can unknowingly spread COVID-19, a fact that needs to be included in epidemic models \cite{Patil:2020}. 
 
Similarly to the work from Saudi Arabia, researchers in Wuhan used the SEIR model to analyze the impacts of public health measures such as quarantines and restrictions of movement \cite{Hou:2020}. Following the time-varying updates recommended in \cite{Chen}, researchers in Portugal dynamically adjusted the exposure rates and other parameters in order to simulate infected asymptomatic individuals who can spread the virus \cite{Teles:2020}. 

Another methodology that has been put to the use in the fight against COVID-19 is agent-based simulation modeling. Even before SARS-CoV-2 first appeared, researchers have been using simulation in conjunction with transit data; the insight is that population movements will critically affect the spread of diseases \cite{Agent-BasedTransportationModels}. As far as COVID-19 is concerned, agent-based simulation models have been used to test the effect of public health mitigation efforts. As an example, in \cite{SILVA2020110088} using agent-based simulation models{\color{black},} the researchers conclude that traditional measures such as mask-wearing and social distancing, as well as lockdowns are viable tools in the fight against COVID-19.

{\color{black}The work presented here is heavily motivated by the literature on diffusion processes on networks (see \cite{colizza2006role}). COVID-19 and its spread is no exception, with many works pointing to the relationship between outbreaks and population movements through the transportation network \cite{jia2020population,doi:10.1126/science.abb4218}. Since 2020, we have seen a multitude of works investigating the network spreading dynamics in air and rail networks as well as public transit \cite{wu2020nowcasting,SUN2021115,li2021assessing,MO2021102893}.}

Finally, we discuss time series models. Autoregressive Integrated Moving Average (ARIMA) models regress a forecast value onto previous values of the time series \cite{HALLETT1986125}. Thus, ARIMA models seek to describe autocorrelations in the time series data \cite{Hyndman:2018}. In India, researchers used ARIMA to model and predict COVID-19 infections \cite{Tandon:2020}, with higher accuracy {\color{black}of} other moving average and exponential smoothing models. Still in India, other research analyzed COVID-19 spreading trends using both an ARIMA and a Holt-Winters model (Holt-Winters accounts for trends and seasonality) \cite{Panda:2020}. The accuracy of the models (during the time period specified) proved very high, at 99.8\%. Another example of ARIMA and Holt-Winters models comes from Jakarta \cite{Sulasikin:EtAl:2020}, finding that ARIMA outperforms the other time series approaches. Last, ensemble methods include a variety of time series models; the final prediction of an ensemble model is a combination of the time series models included \cite{Kotu:2014}. Such an ensemble model was put to use in Nigeria.  The time series model, called Prophet, processed missing values, seasonal effects, and outliers, allowing it to perform well against other models for predicting spread \cite{Abdulmajeed:2020}. 

Researchers employing these techniques across the world can help leaders interdict the spread of the virus. {\color{black} What we mean by this statement is to use spread prediction in a way that informs mobility policy such that threat to human life is minimized. A recent interdiction policy, motivated by COVID-19, is presented in \cite{roy2021effectiveness}. Interestingly, the authors utilize the mobility data and a set of different network science notions on a network obtained from the districts and boroughs in New York City. Outside the context of viral spread and epidemics, researchers have investigated the idea of using betweenness centrality and extensions, such as betweenness-accessibility \cite{sarlas2020betweenness} to identify the most critical links (i.e., streets or main arteries) and nodes (i.e., zip codes, cities, or counties) whose interdiction or closure lead to better isolation of areas. While our work does not focus on interdiction, our contributions can help policy-makers identify parts of the network that are more susceptible to increases in positivity rate.}

\section{The Generalized Network Autoregressive Process}
\label{sec:methods}

In this section, first we describe the Generalized Network Autoregressive Process (GNAR) \cite{GNARreference} and the associated \texttt{R} package \cite{GNARCRAN}. Then, we present the way that we adapt the GNAR model to our problem. We also provide the different metrics that we use to evaluate the performance of time series models.

\subsection{The GNAR Process}

Suppose we have a directed graph $\mathcal{G} = (N, E)$ where $N$ is a set of nodes ($N = \{1, \ldots, n\}$) and $E$ is a set of edges. Suppose we have an edge $e = (i, j) \in E$ for $i, j \in N$ and suppose a direction of $e$ is from a node $i$ to a node $j$, then we write it as $i \to j$.  For any $A \subset N$ we define the {\em neighbor set} of $A$ as follows:
\[
\mathcal{N}(A) := \{j \in N/A| i \to j , \mbox{ for } i \in A\}.
\]
The {\em $r$-th stage neighbors} of a node $i \in N$ is defined as
\[
\mathcal{N}^{(r)}(i) := \mathcal{N}\{\mathcal{N}^{(r-1))}\}/[\{\cup_{q = 1}^{r-1}\mathcal{N}^{(q)}(i)\}\cup \{i\}],
\]
for $r = 2, 3, \ldots$ with $\mathcal{N}^{(1)}(i) = \mathcal{N}(\{i\})$. 

Under this model, we assume that we can assign a weight $\mu_{i, j}$ on an edge $(i, j)$.  We define a distance between nodes $i, j \in N$ such that there exists an edge $(i, j) \in E$ as $d_{i, j} = \mu_{i, j}^{-1}$.  Then we define  
\begin{equation}\label{eq:normalizedwt}
    w_{i, k} =\frac{\mu_{i, k}}{\sum\limits_{l \in \mathcal{N}^{(r)}(i)} \mu_{i, l}}.
\end{equation}

The GNAR model uses a {\em covariate} for an edge affect in different types of nodes by an additional attribute, such as infected or not infected in an epidemiological network. 
Assume that a covariate takes discrete values $\{1, \ldots , C\} \subset \mathbb{Z}$.  Then, let $w_{i, k, c}$ be $w_{i, k}$ for a covariate $c$ such that 
\[
\sum_{q \in \mathcal{N}^{(r)}(i)}\sum_{c = 1}^C w_{i, q, c} = 1.
\]

Now we are ready to define the generalized network autoregressive processes (GNAR) model.  Suppose we have a vector of random variables in 
\[
X_t := (X_{1, t}, \ldots , X_{n, t}) \in \mathbb{R}^n
\]
which varies over the time horizon and each random variable associates with a node.
For each node $i \in N$ and time $t \in \{1, \ldots , T\}$ a generalized network autoregressive processes model of order $(p, [s]) \in \mathbb{N} \times (\mathbb{N}\cup \{0\})^p$ on a vector of random variables $X_t$ is
\begin{equation}\label{eq:gnar}
    X_{i, t}:= \sum_{j = 1}^p \left(\alpha_{i, j}X_{i, t-j} + \sum_{c = 1}^C\sum_{r = 1}^{s_j}\beta_{j, r, c} \sum_{q \in \mathcal{N}_t^{(r)}(i)} w_{i, q, c}^{(t)}X_{q, t-j}\right)
\end{equation}
where $p \in \mathbb{N}$ is the maximum time lag, $[s]:= (s_1, \ldots , s_p)$, $s_j \in \mathbb{N} \cup \{0\}$ is the maximum stage of neighbor dependence for time lag $j$, $\mathcal{N}_t^{(r)}(i)$ is the $r$th stage neighbor set of a node $i$ at time $t$, and  $w_{i, q, c}^{(t)} \in [0, 1]$ is the connection weight between node $i$ and node $q$ at time $t$ if the path corresponds to covariate $c$. $\alpha_{i, j} \in \mathbb{R}$ is a parameter of autoregression at lag $j$ for a node $i \in N$ and $\beta_{j, r, c} \in \mathbb{R}$ corresponds to the effect of the $r$th stage neighbors, at lag $j$,
according to a covariate $c = 1, \ldots , C$.  

\subsection{COVID-19 Analysis Using GNAR}

In order to apply the {GNAR} model defined in this section to the county network on {COVID-19} data, we set variables as follows.  

Note that the {GNAR} model conducts a time series analysis on the time series data on the networks.  The {GNAR} model assumes that the topology of the network is fixed over the time horizon $t > 0$.  In this research, the network is the county network $\mathcal{G} = (N, E)$, where each node $i \in N$ is a county in the particular state in the {US} and we draw an edge $(i, j) \in E$ between a county $i \in N$ and a county $j \in N$ if and only if a county $i$ has commuters traveling to a county $j$. A weight $\mu_{i, j}$ on each edge $(i, j) \in E$ is the number of commuters from a county $i \in N$ and a county $j \in N$. The {GNAR} model assumes that these weights $\mu_{i, j}$ are fixed over the time horizon $t > 0$.  Therefore, the input of the {GNAR} package includes these variables.  Now, {\color{black}what we wish to infer using the {GNAR} model are random variables}
\[
X_t := (X_{1, t}, \ldots , X_{n, t}) \in \mathbb{R}^n,
\]
where $X_{i, n}$ is the number of {COVID-19} cases of deaths from {COVID-19} at a county $i \in N$ at the time $t > 0$.

In this research, we do not have differences between all nodes, i.e., we treat all counties in $N$ as the same type. Therefore, we ignore this covariate index $c$, rendering the formulation of the {GNAR} model as follows. For each county $i \in N$ and time $t \in \{1, \ldots , T\}$ a generalized network autoregressive processes model of order $(p, [s]) \in \mathbb{N} \times (\mathbb{N}\cup \{0\})^p$ on a vector of the numbers of {COVID-19} cases of deaths from {COVID-19} $X_t$ is
\begin{equation}\label{eq:gnar2}
    X_{i, t}:= \sum_{j = 1}^p \left(\alpha_{i, j}X_{i, t-j} + \sum_{r = 1}^{s_j}\beta_{j, r} \sum_{q \in \mathcal{N}_t^{(r)}(i)} w_{i, q}^{(t)}X_{q, t-j}\right)
\end{equation}
where $p \in \mathbb{N}$ is the maximum time lag, $[s]:= (s_1, \ldots , s_p)$, $s_j \in \mathbb{N} \cup \{0\}$ is the maximum stage of neighbor dependence for time lag $j$, $\mathcal{N}_t^{(r)}(i)$ is the $r$th stage neighbor set of a county $i$ at time $t$, and  $w_{i, q}^{(t)} \in [0, 1]$ is the connection weight between a county $i$ and a county $q$ at time $t$. $\alpha_{i, j} \in \mathbb{R}$ is a user specific parameter (tuning parameter) of autoregression at lag $j$ for a county $i \in N$ and a user specific parameter (tuning parameter) $\beta_{j, r} \in \mathbb{R}$ corresponds to the effect of the $r$th stage neighbors, at lag $j$. These user specific parameters are defined by a user. In this paper, we select three combinations of tuning parameters after conducting a model selection discussed in Section \ref{subsection:modelparam}.  

In addition, note that $w_{i, k}$ for a county $i \in N$ and its neighbor $k \in \mathcal{N}^{(r)}_t$ is computed using the equation (\ref{eq:normalizedwt}).

\subsubsection{Model parameters} \label{subsection:modelparam}
The {GNAR} package takes in a number of parameters for its predictive time series models. For both the cases and the deaths, we adjusted two {GNAR} parameters to create three unique models. The first model fit applies a non-negative integer, {\tt{alphaOrder}} = 1, that specifies a maximum time-lag of 1 to model along with a vector of length {\tt{betaOrder}} = 0, which specifies the maximum neighbor set to model at each of the time lags \cite{GNARCRAN}. These parameters represent the time lag, $p$, and the maximum stage of neighbor dependence for each of the time lags, $[s]$, as discussed above. The second model sets ${\tt{alphaOrder}} = 0$ and ${\tt{betaOrder}} = 1$. The third model is the default model in {GNAR}, with no parameter modifications, making both ${\tt{alphaOrder}} = 0, {\tt{betaOrder}} = 0$. We conduct a model selection by changing {\tt{alphaOrder}} and {\tt{betaOrder}} from $0$ to $5$ independently.  Table \ref{table:modelparam} provides a summary of the model parameter combinations.

\begin{table}
\centering
\caption[Model Parameter Summary]{Model Parameter Summary. We vary the value of {\tt{alphaOrder}} and {\tt{betaOrder}} to create three different models for the {COVID-19} cases and deaths.}
\begin{tabular}{|c|c|c|}
\hline
\textbf{Model 1}                                                       & \textbf{Model 2}                                                       & \textbf{Model 3}                                                       \\ \hline
\begin{tabular}[c]{@{}c@{}}alphaOrder = 1\\ betaOrder = 1\end{tabular} & \begin{tabular}[c]{@{}c@{}}alphaOrder = 0\\ betaOrder = 1\end{tabular} & \begin{tabular}[c]{@{}c@{}}alphaOrder = 0\\ betaOrder = 0\end{tabular} \\ \hline
\end{tabular}
\label{table:modelparam}
\end{table}

Because there are two prediction options (cases and deaths) and three model parameter selections, in total, we create 6 different combinations (e.g., Deaths - Model 1). Moreover, since we predict by state, each state has these $6$ models for comparison.

\subsection{Evaluation performance}
\label{subsection:performance}

Measuring performance in traditional statistics often calls for measures of performance such as RMSE and adjusted $R^2$. Although easily calculated, these measures {\color{black}do} not measure errors in terms of the time horizon \cite{Guo:2012}. For outputs of a predictive time series model, performance can be measured by the 
mean absolute percentage error (MAPE) and 
the mean absolute scaled error (MASE). The MAPE measures an estimated average of a model's forecast performance over the time horizon, while the MASE measures the ratio of an estimated absolute error of the forecast divided and estimated absolute error of the na\"ive forecast method over the time horizon \cite{hyndman2006another}. The MAPE commonly falls between 0 and 1, but can be skewed outside this range if actual values are close to zero \cite{hyndman2006another}. The {\color{black}MASE} is less than one if a model has smaller error than the na\"ive model's error and if a model has greater error than the na\"ive model's error, then it is greater than one \cite{hyndman2006another}. The MAPE and MASE are defined by the following equations:

\begin{equation}
MAPE = \frac{\sum_{t=1}^{N}|\frac{Y_t-F_t}{Y_t}|}{N},
\end{equation}
\begin{equation}
MASE = \frac{\sum_{t=1}^{N}|\frac{Y_t-F_t}{Y_t-Y_{t-1}}|}{N},
\end{equation}
where $Y_t$ is the observation at time $t$, $F_t$ is the predicted value, and $Y_t-Y_{t-1}$ is error of the one-step na\"ive forecast.

In order for a model to have predictive power, its MASE must exceed the accuracy of the respective na\"ive model and we say it has a good forecasting if a model has the MAPE less than $0.2$ \cite{Lewis:1982}.

When we measure {\color{black}the} performance of each model in terms of the MASE and MAPE, we apply a {\em rolling horizon design} \cite{10.2307/2630927} in this paper.  A rolling horizon design for a time series model is to assess accuracy of a time series model such that it updates the forecasted value successively using different subsets of previous and current observations, and then it takes averages of the performance of the model for different time periods.

\section{Data}
\label{sec:data}

We obtain the data for this work from the United States Census Bureau (USCB), the United States of America Facts (USAFacts), and the Center for Disease Control and Prevention (CDC). The {USAFacts} obtains their data from the {CDC} \cite{CDC:2020} and updates the daily death count on their website \cite{AboutUSA15}. Manipulating this data in {\tt{Python}}, we transform the data into a usable format for the {GNAR} package, create our models, and assess them using a variety of evaluation performance metrics. 

\subsection{Data description and limitations} \label{subsection:data_description}

The data is entirely numerical, with no categorical predictors or response variables. No transformations are applied to the original data for the proposed models. We also note that the {COVID-19} cases and deaths data meet the assumption of stationarity of errors because the noise of the data does not depend on the time at which the data was observed \cite{Hyndman:2018}. Autoregressive models require stationarity of the errors, meaning that the series' variance must be constant over a long time period \cite{SS:2021}. 

Furthermore, we assume that the {COVID-19} data is complete and accurate. Although human error and reporting standards affect the number of deaths and cases sometimes, on any given day, we assume the data obtained from the CDC is accurate. Additionally, the data is autoregressive. Last, we assume the presence of no outliers \cite{Frost:2020}.
The principal limitation of this data involves the constant nature of the commuting network structure \cite{USCB:2019}. The {USCB} compiled this commuting data over a five year period from 2011 to 2015, giving it a static property. We thus assume that the traffic and commuting patterns by county remain the same through the time of the {COVID-19} pandemic.


\subsection{County network}

The United States comprises 3,143 county or county equivalents as of 2020 \cite{USCB:2020a}. 48 states use the term ``county'' to describe their administrative districts while Louisiana and Alaska use the terms ``parishes'' and ``boroughs,'' respectively. Each county is assigned a unique five-digit {FIPS} code. The first two digits represent the state's {FIPS} code, while the latter three digits represent the county's {FIPS} code within the state. This number serves as a uniform index for each county, facilitating county data sorting and filtering. 

The number of counties per a state varies widely across the United States, regardless of a state's geographic size, population, or terrain. For example, Rhode Island, the state with the smallest land area, has 5 counties, while Alaska, the state with the largest land area, has 29 counties \cite{USCB:2020b}. However, Delaware contains the least amount of counties and Texas contains the most. Moreover, the majority of the country's population lives in only 143 of the 3143 counties as of 2020. Table \ref{table:countynumbers} {\color{black} and Fig.~\ref{fig:uscounties} display} the number of counties in each state. 

\begin{table}
\centering
\caption[Number of counties by state]{A list of all United States counties per state (including the District of Columbia). As can be seen, the number of counties per state varies widely. Source: \cite{TFI:2021}.}
\begin{tabular}{|l|c||l|c|}
\hline
\textbf{Sate}        & \textbf{Counties} & \textbf{State} & \textbf{Counties} \\ \hline
Alabama              & 67                & Montana        & 56                \\ \hline
Alaska               & 29                & Nebraska       & 93                \\ \hline
Arizona              & 15                & Nevada         & 17                \\ \hline
Arkansas             & 75                & New Hampshire  & 10                \\ \hline
California           & 58                & New Jersey     & 21                \\ \hline
Colorado             & 64                & New Mexico     & 33                \\ \hline
Connecticut           & 8                 & New York       & 62                \\ \hline
Delaware             & 3                 & North Carolina & 100               \\ \hline
District of Columbia & 1                 & North Dakota   & 53                \\ \hline
Florida              & 67                & Ohio           & 88                \\ \hline
Georgia              & 159               & Oklahoma       & 77                \\ \hline
Hawaii               & 5                 & Oregon         & 36                \\ \hline
Idaho                & 44                & Pennsylvania   & 67                \\ \hline
Illinois             & 102               & Rhode Island   & 5                 \\ \hline
Indiana              & 92                & South Carolina & 46                \\ \hline
Iowa                 & 99                & South Dakota   & 66                \\ \hline
Kansas               & 105               & Tennessee      & 95                \\ \hline
Kentucky             & 120               & Texas          & 254               \\ \hline
Louisiana            & 64                & Utah           & 29                \\ \hline
Maine                & 16                & Vermont        & 14                \\ \hline
Maryland             & 24                & Virginia       & 133               \\ \hline
Massachusetts        & 14                & Washington     & 39                \\ \hline
Michigan             & 83                & West Virginia  & 55                \\ \hline
Minnesota            & 87                & Wisconsin      & 72                \\ \hline
Mississippi          & 82                & Wyoming        & 23                \\ \hline
Missouri             & 115               &                &                   \\ \hline
\end{tabular}
\label{table:countynumbers}
\end{table}

{\color{black}
\begin{figure}
    \centering
    \includegraphics[width=\textwidth]{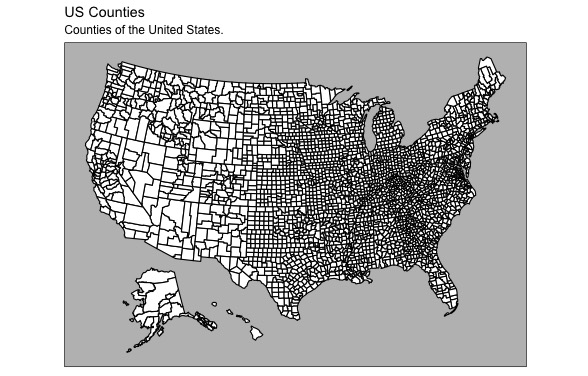}
    \caption{All counties in the US plotted by the {\tt ggplot2} \cite{ggplot2} and {\tt usmap} \cite{usmap} packages in {\tt R}.}
    \label{fig:uscounties}
\end{figure}
}

Additionally, it is not uncommon for county information to change as time goes on. Counties can divide, merge, or rename themselves at any time, even if that time does not fall on a census year. For example, Colorado created Broomfield County from merging parts of other counties \cite{BC:2021}. Shannon County, South Dakota renamed itself to Ogala County in 2015 out of respect to its Native American heritage \cite{ShannonCounty}. Of course, when new counties are formed or renamed, they are assigned new FIPS codes, which can complicate reporting of statistics later on. Regardless, many counties and states have protocols in place to prevent such mistakes.

In this work, we construct a {network structure} using the original commuting data from the {USCB} using {\tt{Python}}. The commuting data comes in the form of a data frame with three columns: the county from which the individuals commute, the county to which the individuals commute, and the number of commuters. This represents a flow structure, where we can deduce how many commuters commute from one county to the next. 
Using {\tt{Python}}, we transform this flow structure into a matrix format, with the row and column entries representing the ``From'' and ``To'' columns of the original data. Thus, one can easily search in this new commuting data matrix for how many commuters go from one county to the next. 

In this research we divide the county network in each state.
We designed the county commuting network structure for each state with the following information:
\begin{itemize}
    \item Workers commuting from within-state counties to within-state counties.
    \item Workers commuting from out-of-state counties to within-state counties.
    \item Workers commuting from within-state counties to out-of-state counties.
\end{itemize}
Dividing the network into states allows us to see more localized trends in {COVID-19}, instead of considering the entire country at once. States can act as ``communities'' in the country's commuting network. Communities in network science are groups of nodes with similar characteristics \cite{newmanbarabasi}.

\section{Computational experiments}
\label{sec:results}

As discussed earlier, GNAR takes a univariate time series dataset along with an underlying network structure in order to create a predictive time series model. After transforming the data for the {GNAR} model, we fit three models for each prediction type (for COVID-19 cases and COVID-19 deaths), giving us 6 models for each state. We will be selecting 5 states to showcase the results on; since the 5 states will be tested on 6 models each, this givs us a total of 30 individual models. We can evaluate the models for prediction accuracy graphically, using the {MAPE} and {MASE} as measures of performance.

In order to determine if our models would perform in a similar fashion across different states, we select a diverse array of states based on vaccination rates. Vaccination rates could affect the time-varying number of cases and deaths in a state, potentially leading to differing model performances. Hence, we choose the following states for our analysis: Rhode Island, Massachusetts, California, Florida, and Arkansas. Table \ref{table:vaccinations} describes the vaccination rate (\% of population) and corresponding rank out of 50 of each state we choose.

\begin{table}[h!]
\centering
\caption[State Vaccination Rates]{State vaccination rates as of 11/2021. {\color{black} The vaccination rates capture the percentage of population that is considered fully vaccinated (two doses of the appropriate vaccines, or one dose of a single dose vaccine). The rates help us determine which states to choose for our comparison.} Adapted from \cite{Adams:2021}.}
\begin{tabular}{|c|c|c|}
\hline
\textbf{State} & \textbf{\begin{tabular}[c]{@{}c@{}}Vaccination Rate \\ (\% Population)\end{tabular}} & \textbf{Rank} \\ \hline
Rhode Island   & 71.0                                                                                 & 2             \\ \hline
Massachusetts  & 69.8                                                                                 & 5             \\ \hline
California     & 61.4                                                                                 & 16            \\ \hline
Florida        & 59.8                                                                                 & 22            \\ \hline
Arkansas       & 48.1                                                                                 & 43            \\ \hline
\end{tabular}
\label{table:vaccinations}
\end{table}

We calculate the {MASE} and {MAPE} for each {GNAR} model with respect to each test period within the 40-week forecast. We then calculate the mean, median, and variance of these values. 
In order to determine if a transformation of measurements in a given data helps {\color{black}in} improving {\color{black}the} performance of a model, we test transforming measurements in each county commuting network in the following ways: {\color{black} \begin{enumerate}
    \item logarithm transformation,
    \item square-root transformation, and
    \item normalization.
\end{enumerate}  In our experiments, however, all the above transformations result in minor changes in the performance of the model; therefore, we report here the results using {\color{black}the} original scale of measurements.}

The na\"ive models across all states started with a high {MAPE} at the beginning in the time horizon, sometimes doubling the {MAPE} of all the other models. Additionally, the {MAPE} for {\color{black}the} na\"ive model appears to increase slightly across all states in the last four weeks in the time horizon.  {\color{black}Poor performance of the na\"ive model in the beginning of the time horizon may be caused by a sudden increase of cases and a lack of information in the beginning of the time horizon. }

As we will see below, the {MASE} for Models 1 and 2 in each state exhibit a bimodal ``hump'', with the largest hump centered around week 80 of our dataset. This hump then shows a sharp decrease for the last five weeks of the testing period for all models. Models 1 and 2 perform worse than the baseline na\"ive model during this bimodal hump period. {\color{black}The 80th week mark falls near the end of August of 2021: during that period, everywhere in the US, the number of cases increased at a much slower pace than earlier. The timing coincides with when many people were fully vaccinated (for clarity, full vaccination in our work refers to a two-dose regimen or a single dose of an approved vaccine \cite{cdc:full_vaccination}). Hence, these big ``humps'' in the model performance could be caused by the effect of these vaccinations on the number of cases. Models 1 and 2 are the ones impacted by this increased number of vaccinations. This is because Model 1 and Model 2 are most affected by the numbers of cases in neighboring counties, and vaccination rates in neighbor counties are not necessarily the best predictors for the number of cases in each individual county.}
Overall, {\color{black}since Model 3 seems largely unaffected by correlations between numbers of fully vaccinated individuals in neighbor counties,} Model 3 performs the best, since it outperforms the na\"ive model at almost all times during the time horizon. Earlier, we mentioned that full vaccination refers to individuals that have received two doses or a single dose of the appropriate vaccines. {\color{black}It would be interesting, given updated information and data, to study whether this effect changes when considering an individual as full vaccinated after having received the appropriate booster shot, also referred to as ``up-to-date'' individuals \cite{cdc:full_vaccination}.}

\begin{figure}
\centering
\begin{subfigure}[t]{0.45\textwidth}
\includegraphics[width=\textwidth]{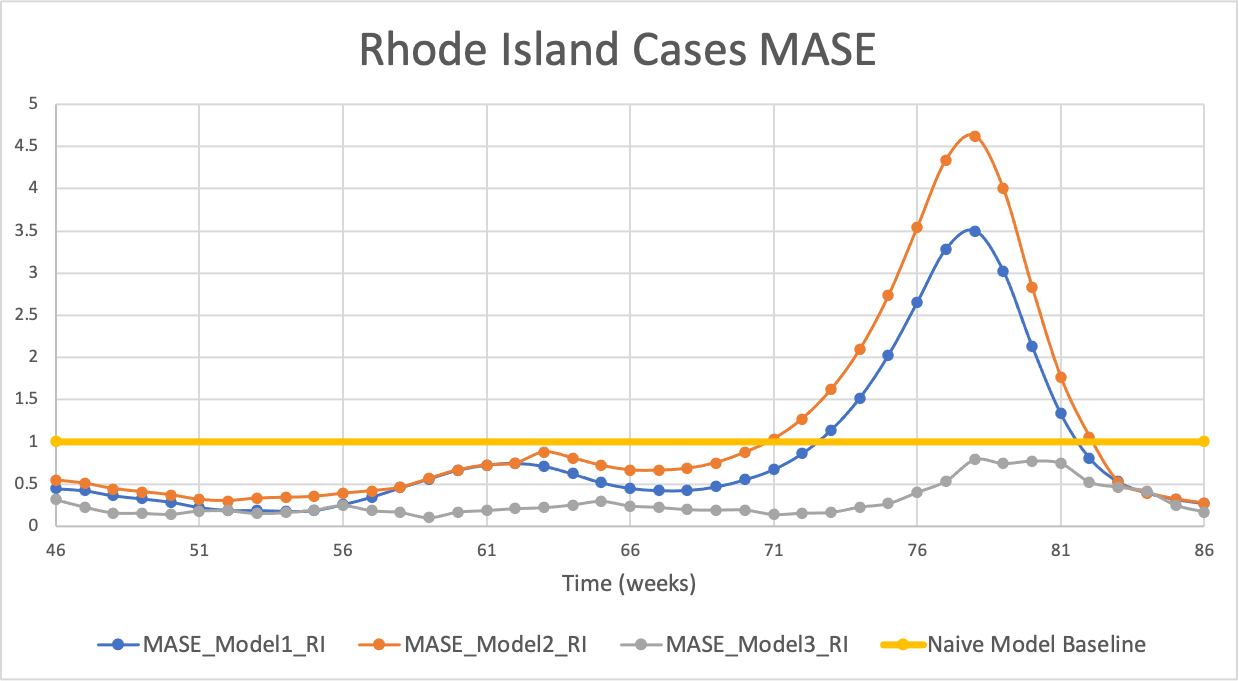}
\caption{Rhode Island Cases {MASE}. The vertical axis represents MASE. Models 1 and 2 exhibit a dramatic increase in {MASE} at approximately week 71, then a sharp decrease at week 77.}
\end{subfigure}\hfill
\begin{subfigure}[t]{0.45\textwidth}
\includegraphics[width=\textwidth]{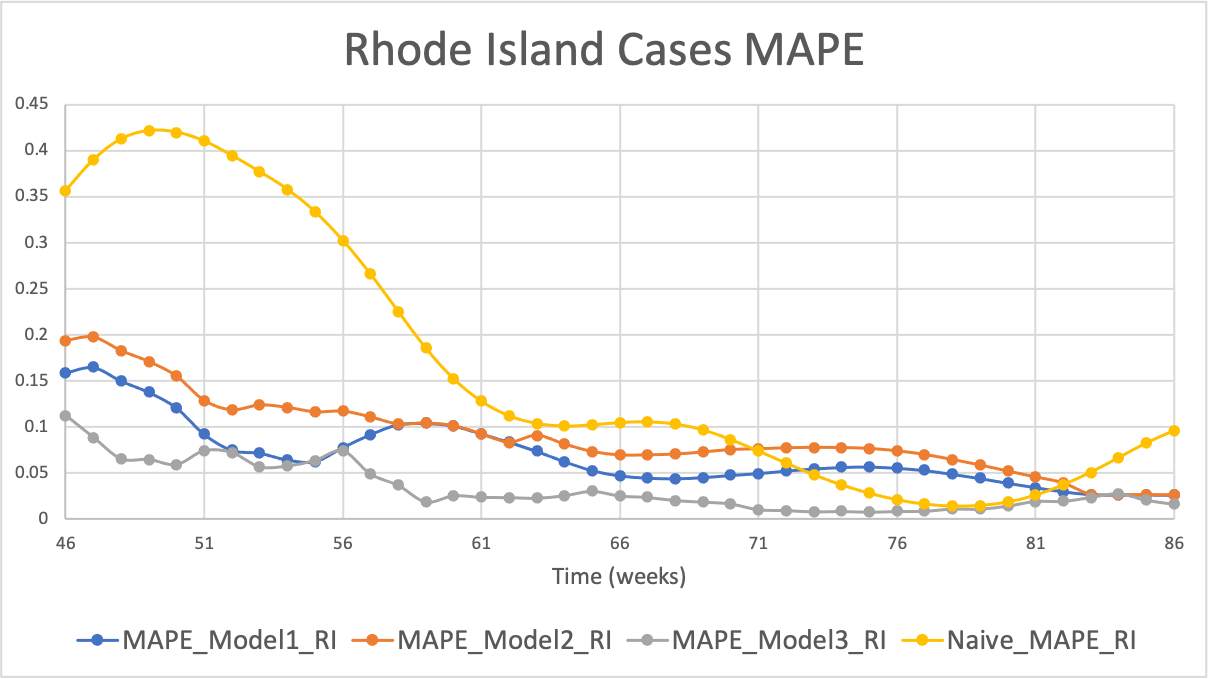}
\caption{Rhode Island Cases {MAPE}. The vertical axis represents MAPE. Models 1 and 2 perform well against the na\"ive model until approximately week 71. Models 1, 2, and 3 all trend downward gradually.}
\end{subfigure}
\begin{subfigure}[t]{0.45\textwidth}
\includegraphics[width=\textwidth]{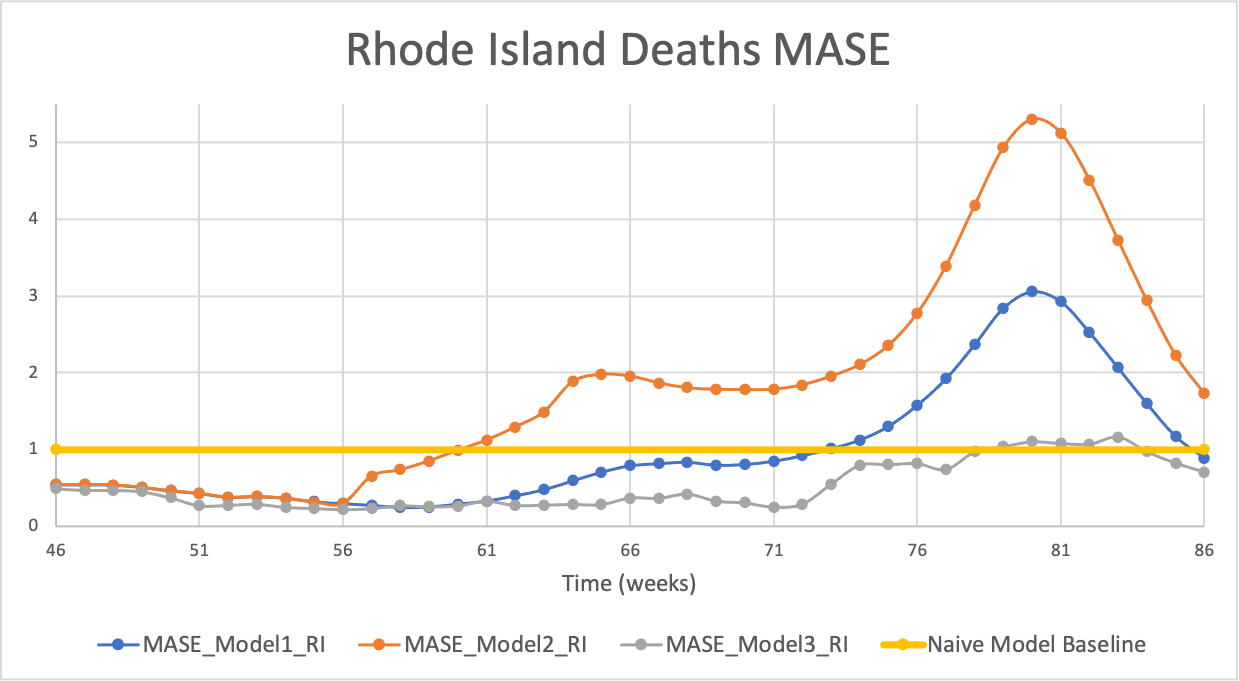}
\caption{Rhode Island Deaths {MASE}. The vertical axis represents MASE. Model 2 starts to perform worse than the na\"ive model at approximately week 61 and continues until the end of the testing period. Model 1 starts to perform worse than the na\"ive model at approximately week 73 and continues until approximately week 85. Model 3 did not perform better than the na\"ive model for about six weeks from approximately week 78 to week 83.}
\end{subfigure}\hfill
\begin{subfigure}[t]{0.45\textwidth}
\includegraphics[width=\textwidth]{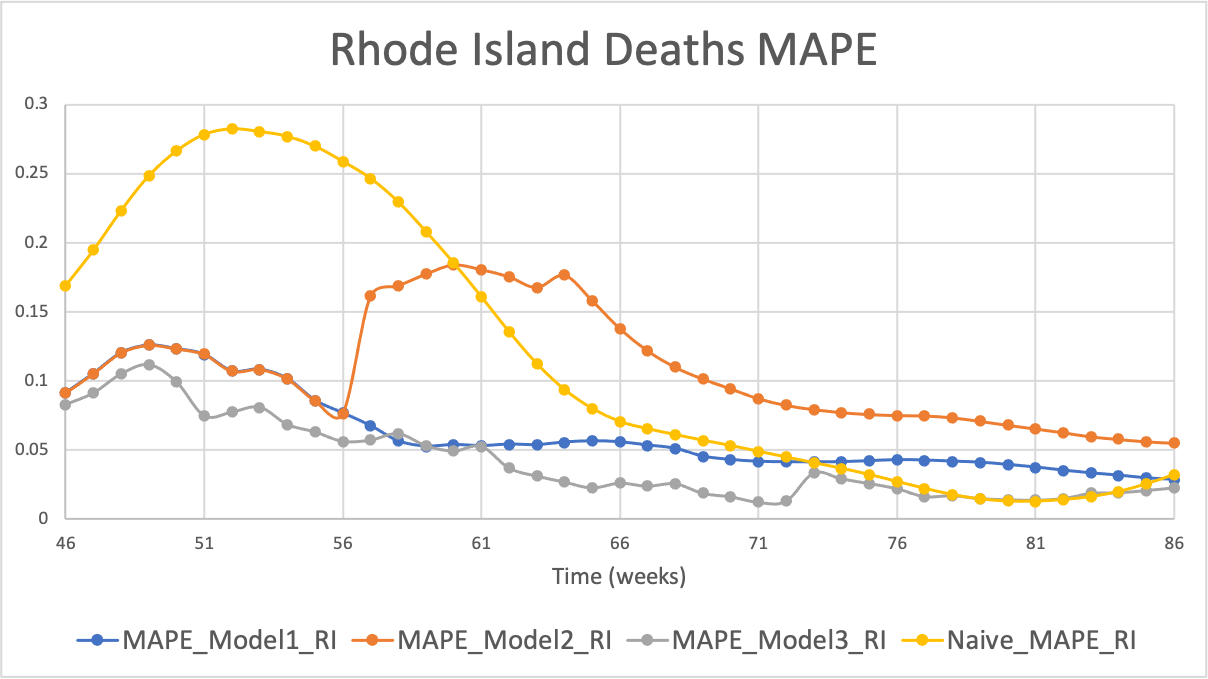}
\caption{Rhode Island Deaths {MAPE}. The vertical axis represents MAPE. Model 2 experienced a sharp increase in {MAPE} at approximately week 56, trending to perform worse than the na\"ive model until the end of the testing period. Model 2 and Model 3 stayed at a lower threshold, with Model 3 outperforming the na\"ive 100\% of the time. }
\end{subfigure}
\caption{The horizontal axis depicts the number of weeks since data collection began on January 22, 2020 for all plots in this figure. The four subfigures discuss the three models for the cases and deaths and the MAPE and MASE evaluation metrics for the state of Rhode Island. Adapted from \cite{USCB:2019} and \cite{USAFacts:2021a}.
}
\label{fig:RI}
\end{figure}

\begin{figure}
\centering
\begin{subfigure}[t]{0.45\textwidth}
\includegraphics[width=\textwidth]{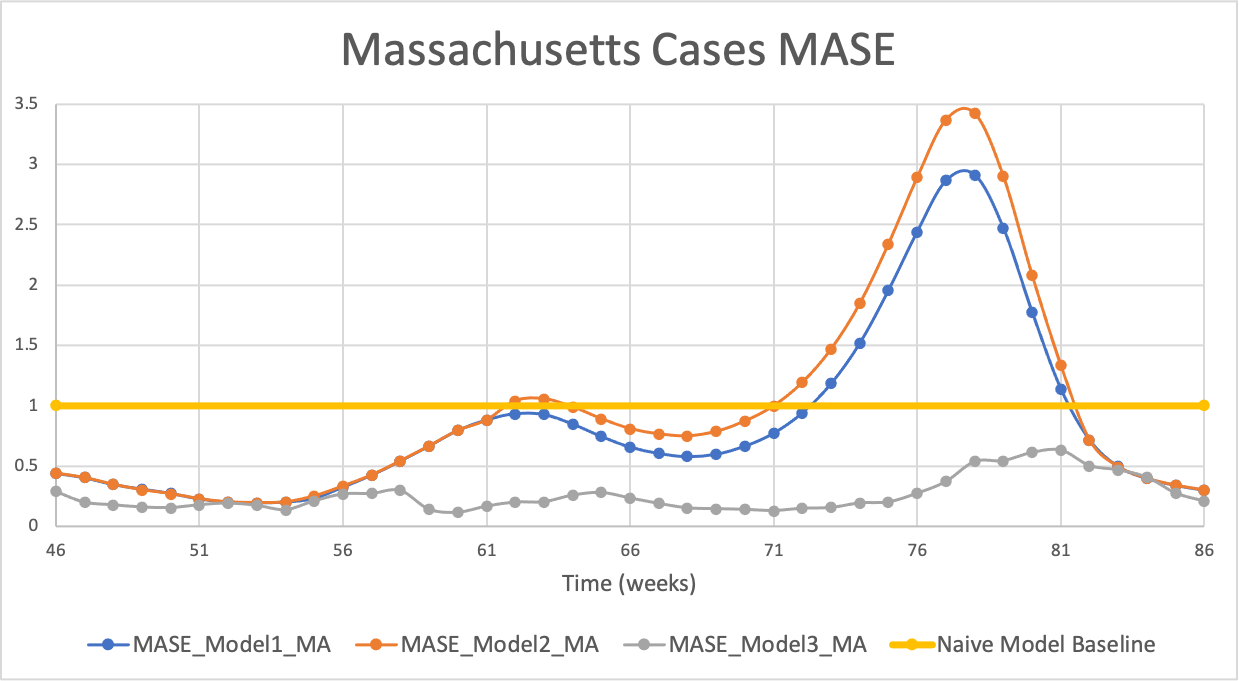}
\caption{Massachusetts Cases {MASE}. The vertical axis represents MASE. Models 1 and 2 exhibit a dramatic increase in {MASE} at approximately week 71, then a sharp decrease at week 78. Models 1 and 2 perform worse than the na\"ive model in this time.}
\end{subfigure}\hfill
\begin{subfigure}[t]{0.45\textwidth}
\includegraphics[width=\textwidth]{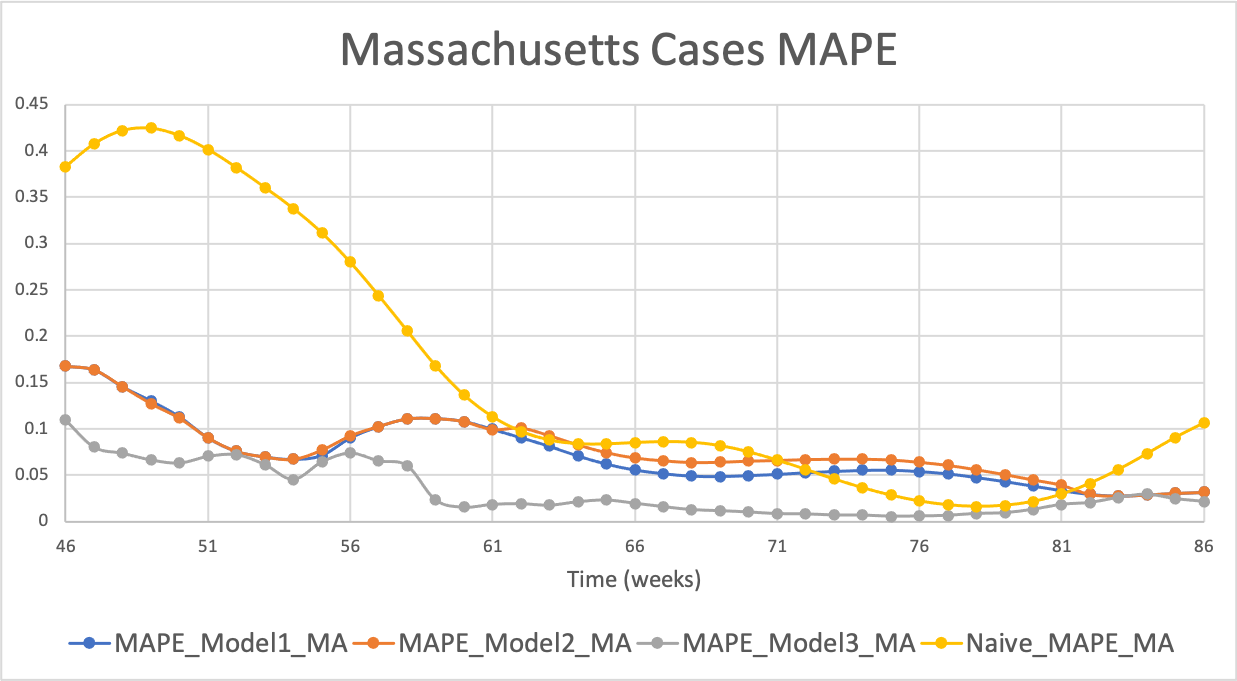}
\caption{Massachusetts Cases {MAPE}. The vertical axis represents MAPE. Models 1 and 2 follow each other closely, performing worse than the na\"ive model from approximately week 71 to week 81. Models 1, 2, and 3 all trend downward gradually.}
\end{subfigure}
\begin{subfigure}[t]{0.45\textwidth}
\includegraphics[width=\textwidth]{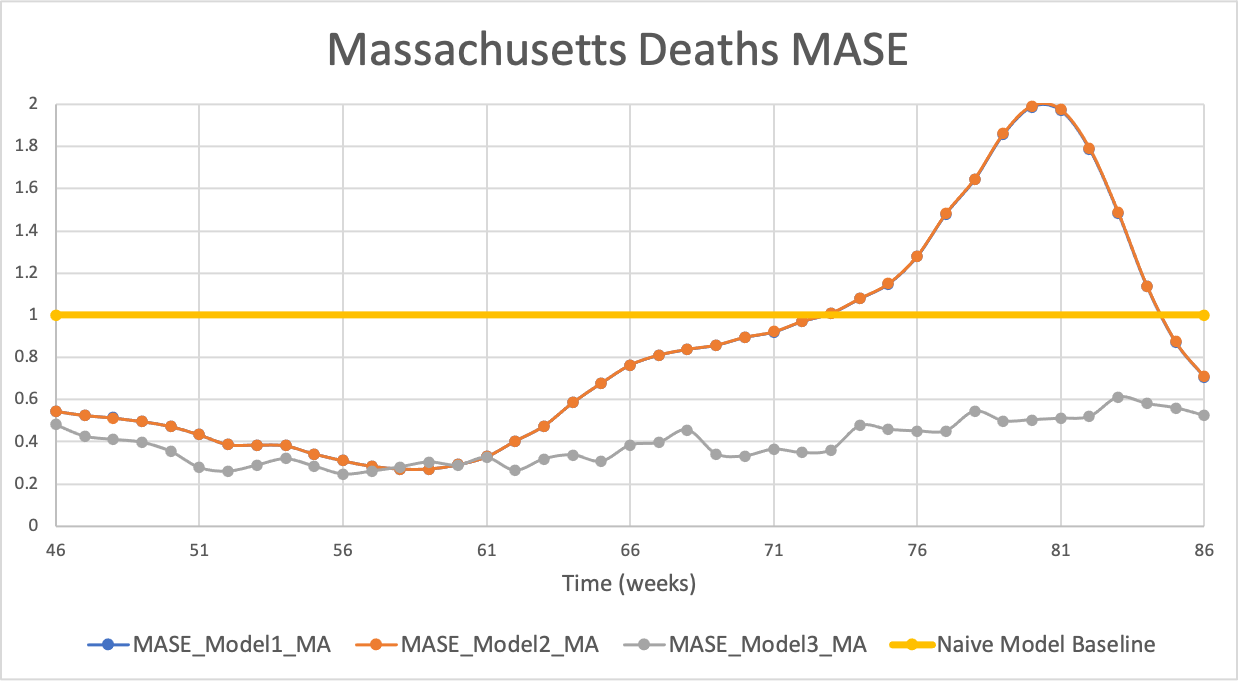}
\caption{Massachusetts Deaths {MASE}. The vertical axis represents MASE. Models 1 and 2 again follow each other very closely, almost overwriting each other. Models 1 and 2 start to perform worse than the na\"ive model at approximately week 72 but end up performing better than it at approximately week 84.}
\end{subfigure}\hfill
\begin{subfigure}[t]{0.45\textwidth}
\includegraphics[width=\textwidth]{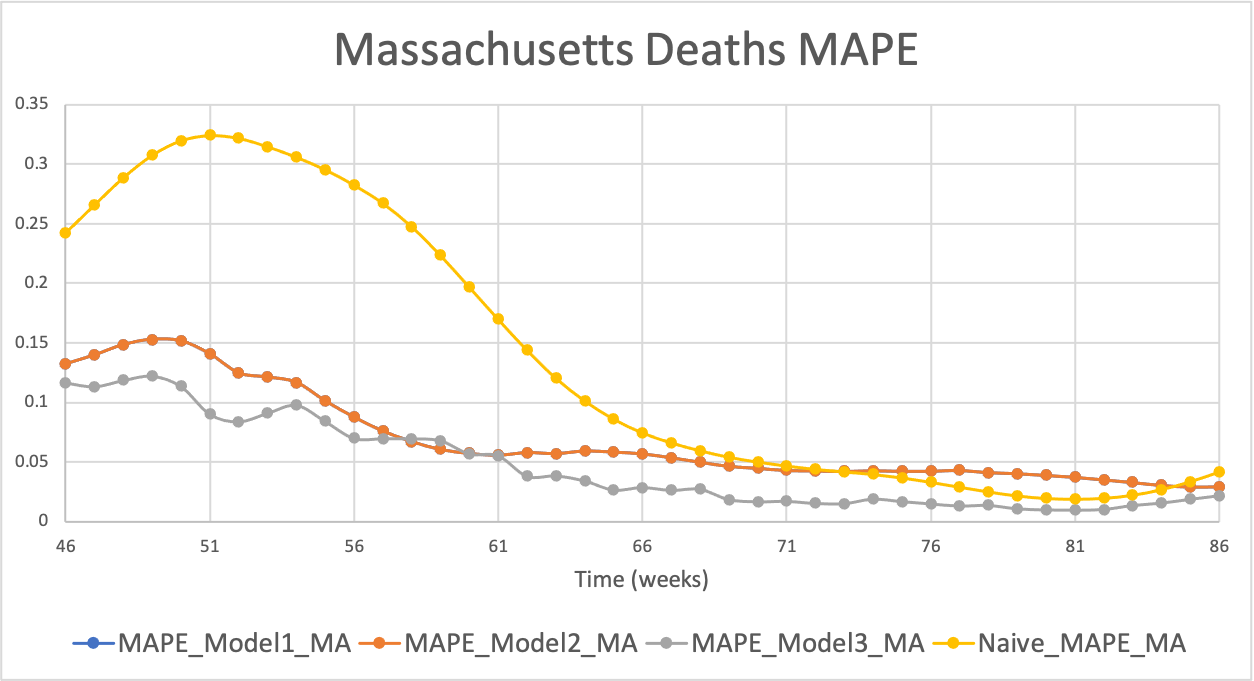}
\caption{Massachusetts Deaths {MAPE}. The vertical axis represents MAPE. Models 1 and 2 follow each other closely, almost overwriting each other. They perform worse than the na\"ive model from approximately week 75 to week 84. All models trend gradually downwards.}
\end{subfigure}
\caption{The horizontal axis depicts the number of weeks since data collection began on January 22, 2020 for all plots in this figure. The four subfigures discuss the three models for the cases and deaths and the MAPE and MASE evaluation metrics for the state of Massachusetts. Adapted from \cite{USCB:2019} and \cite{USAFacts:2021a}.
}
\label{fig:MA}
\end{figure}

\begin{figure}
\centering
\begin{subfigure}[t]{0.45\textwidth}
\includegraphics[width=\textwidth]{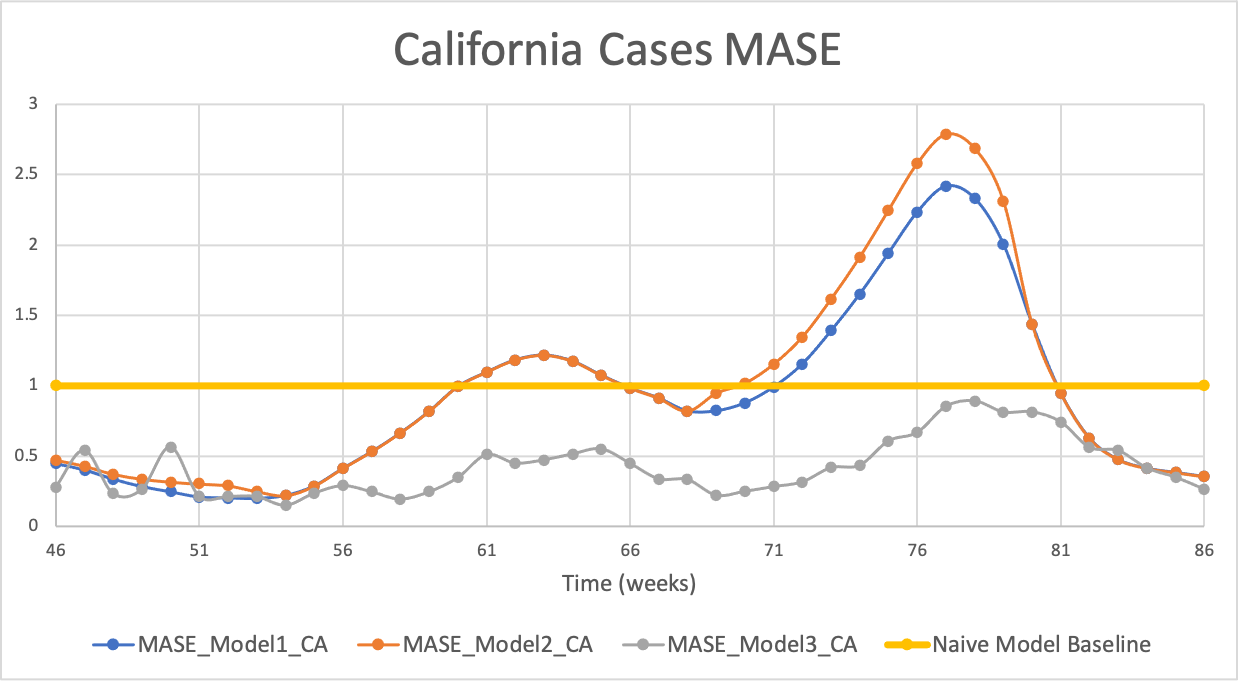}
\caption{California Cases {MASE}. The vertical axis represents MASE. Models 1 and 2 follow each other closely and exhibit a dramatic increase in {MASE} at approximately week 71, then a sharp decrease at week 77. Models 1 and 2 perform worse than the na\"ive model in this time.}
\end{subfigure}\hfill
\begin{subfigure}[t]{0.45\textwidth}
\includegraphics[width=\textwidth]{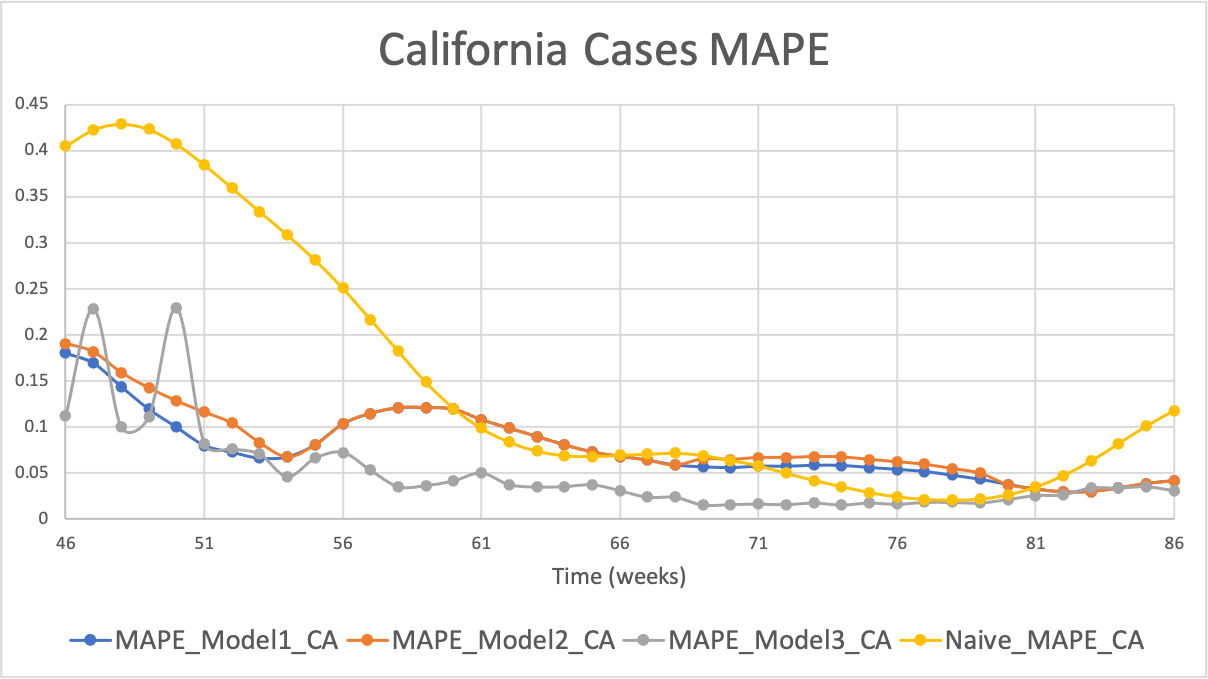}
\caption{California Cases {MAPE}. The vertical axis represents MAPE. Models 1 and 2 follow each other closely, performing worse than the na\"ive model from approximately week 61 to week 80. Models 1, 2, and 3 all trend downward gradually.}
\end{subfigure}
\begin{subfigure}[t]{0.45\textwidth}
\includegraphics[width=\textwidth]{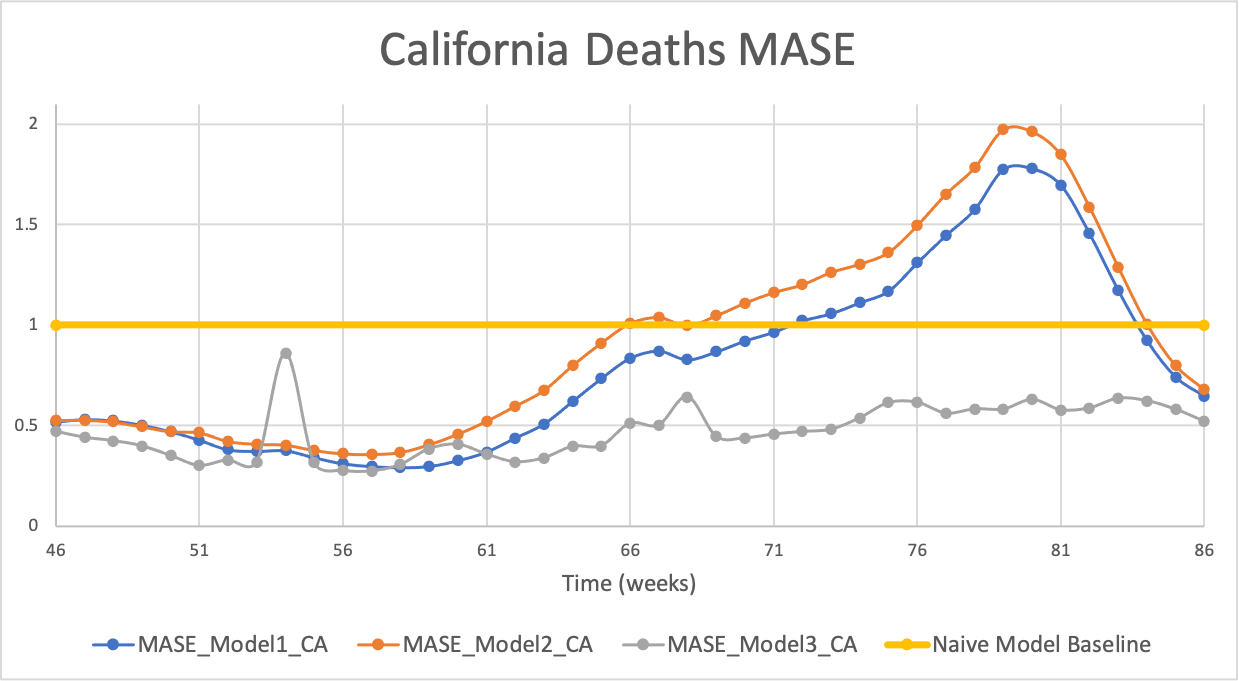}
\caption{California Deaths {MASE}. The vertical axis represents MASE. Models 1 and 2 again follow each other very closely, almost overwriting each other. Models 1 and 2 start to perform worse than the na\"ive model at approximately week 72 but end up performing better than it at approximately week 84.}
\end{subfigure}\hfill
\begin{subfigure}[t]{0.45\textwidth}
\includegraphics[width=\textwidth]{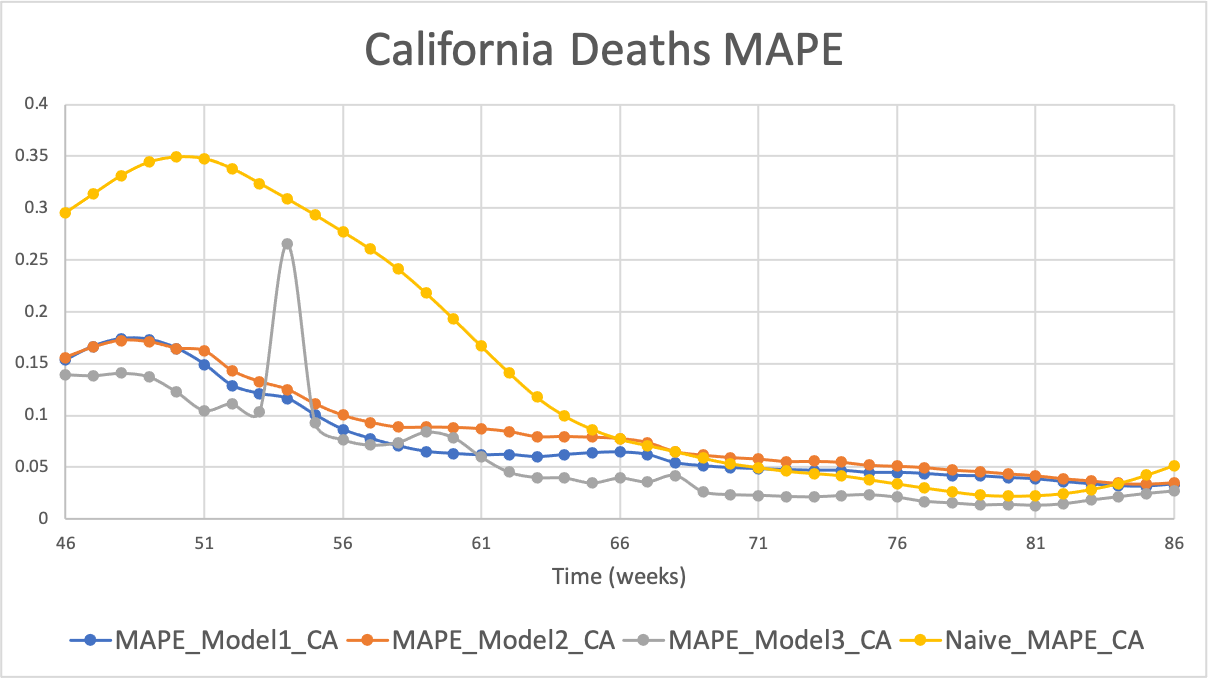}
\caption{California Deaths {MAPE}. The vertical axis represents MAPE. Models 1 and 2 follow each other closely. They perform worse than the na\"ive model from approximately week 72 to week 84. All models trend gradually downwards.}
\end{subfigure}
\caption{The horizontal axis depicts the number of weeks since data collection began on January 22, 2020 for all plots in this figure. The four subfigures discuss the three models for the cases and deaths and the MAPE and MASE evaluation metrics for the state of California. Adapted from \cite{USCB:2019} and \cite{USAFacts:2021a}.
}
\label{fig:CA}
\end{figure}

\begin{figure}
\centering
\begin{subfigure}[t]{0.45\textwidth}
\includegraphics[width=\textwidth]{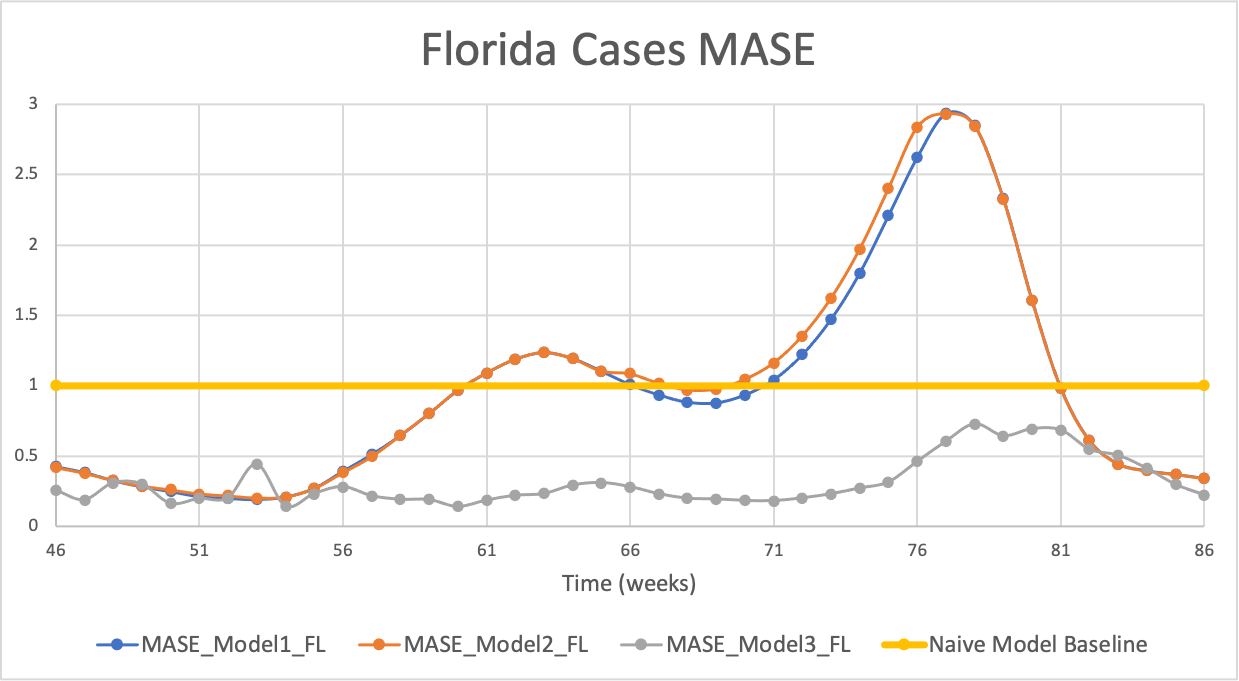}
\caption{Florida Cases {MASE}. The vertical axis represents MASE. Models 1 and 2 follow each other very closely and exhibit a dramatic increase in {MASE} at approximately week 71, then a sharp decrease at week 77. Models 1 and 2 perform worse than the na\"ive model in this time.}
\end{subfigure}\hfill
\begin{subfigure}[t]{0.45\textwidth}
\includegraphics[width=\textwidth]{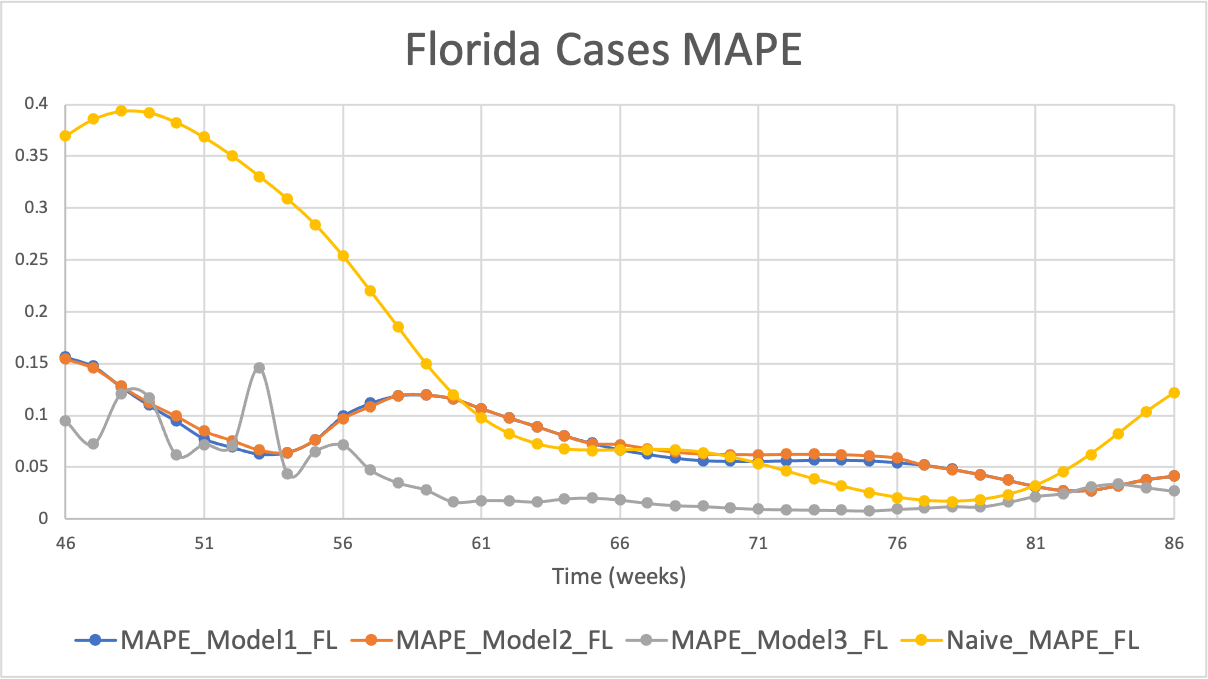}
\caption{Florida Cases {MAPE}. The vertical axis represents MAPE. Models 1 and 2 follow each other closely, performing worse than the na\"ive model from approximately week 61 to week 65 and week 71 to week 81. Models 1, 2, and 3 all trend downward gradually.}
\end{subfigure}
\begin{subfigure}[t]{0.45\textwidth}
\includegraphics[width=\textwidth]{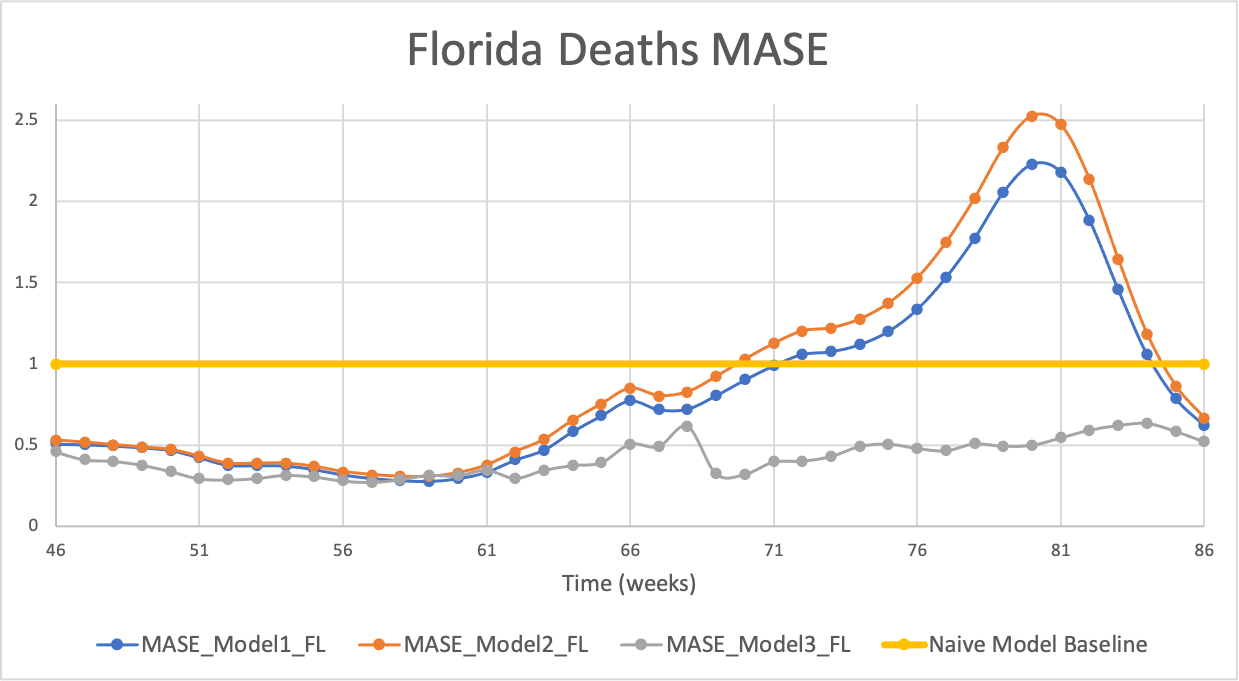}
\caption{Florida Deaths {MASE}. The vertical axis represents MASE. Models 1 and 2 follow each other closely. Models 1 and 2 start to perform worse than the na\"ive model at approximately week 71 but end up performing better than it at approximately week 84.}
\end{subfigure}\hfill
\begin{subfigure}[t]{0.45\textwidth}
\includegraphics[width=\textwidth]{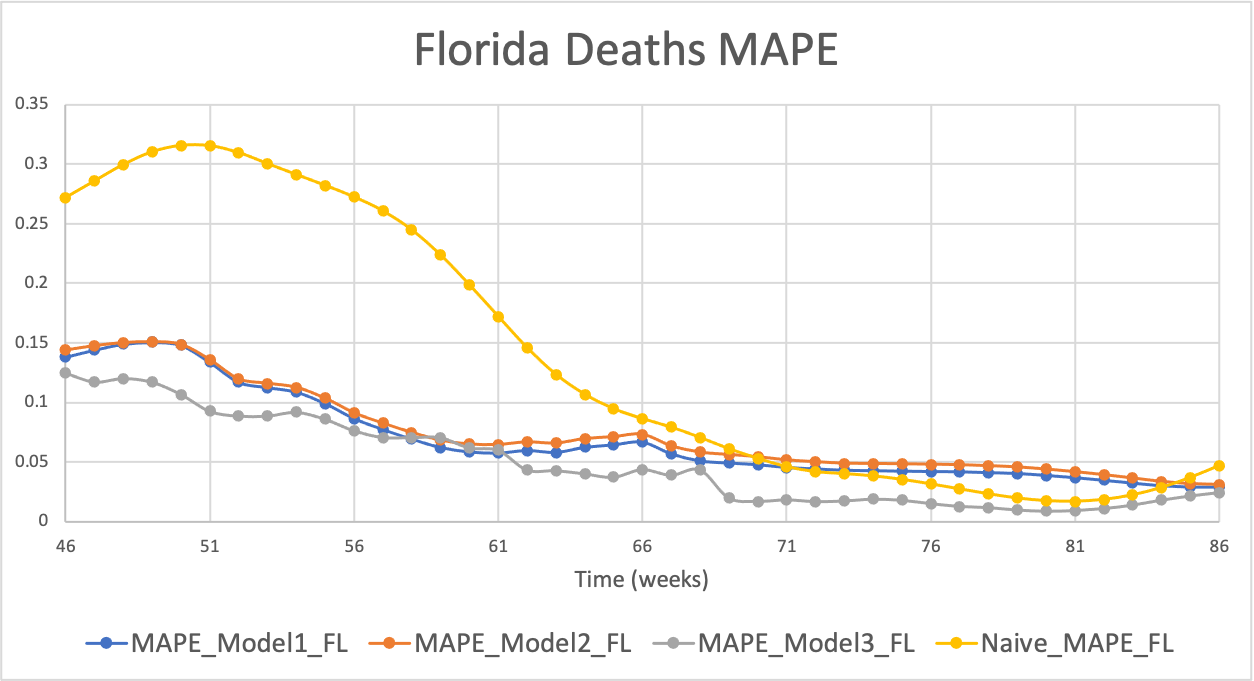}
\caption{Florida Deaths {MAPE}. The vertical axis represents MAPE. Models 1 and 2 follow each other closely. They perform worse than the na\"ive model from approximately week 71 to week 84. All models trend gradually downwards.}
\end{subfigure}
\caption{The horizontal axis depicts the number of weeks since data collection began on January 22, 2020 for all plots in this figure. The four subfigures discuss the three models for the cases and deaths and the MAPE and MASE evaluation metrics for the state of Florida. Adapted from \cite{USCB:2019} and \cite{USAFacts:2021a}.
}
\label{fig:FL}
\end{figure}

\begin{figure}
\centering
\begin{subfigure}[t]{0.45\textwidth}
\includegraphics[width=\textwidth]{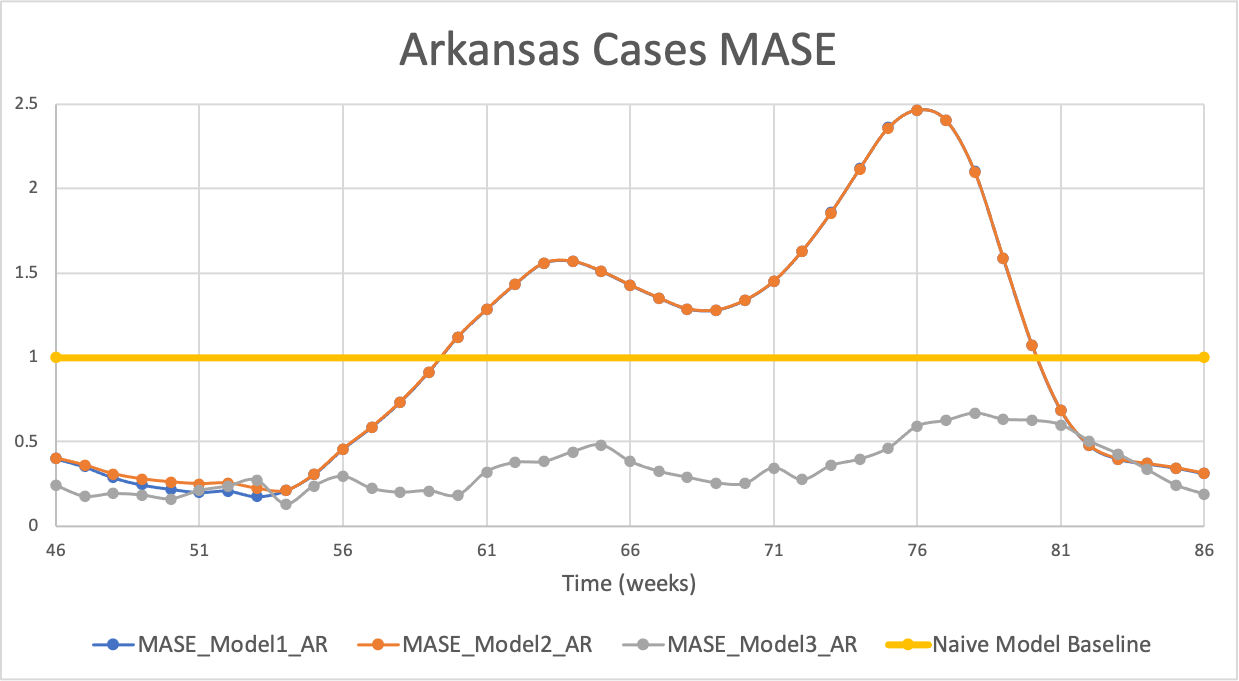}
\caption{Arkansas Cases {MASE}. The vertical axis represents MASE. Models 1 and 2 follow each other very closely and exhibit a dramatic increase in {MASE} much earlier than the other states, at approximately week 56, then a sharp decrease at week 77. Models 1 and 2 perform worse than the na\"ive model in this time but end up performing better by week 80.}
\end{subfigure}\hfill
\begin{subfigure}[t]{0.45\textwidth}
\includegraphics[width=\textwidth]{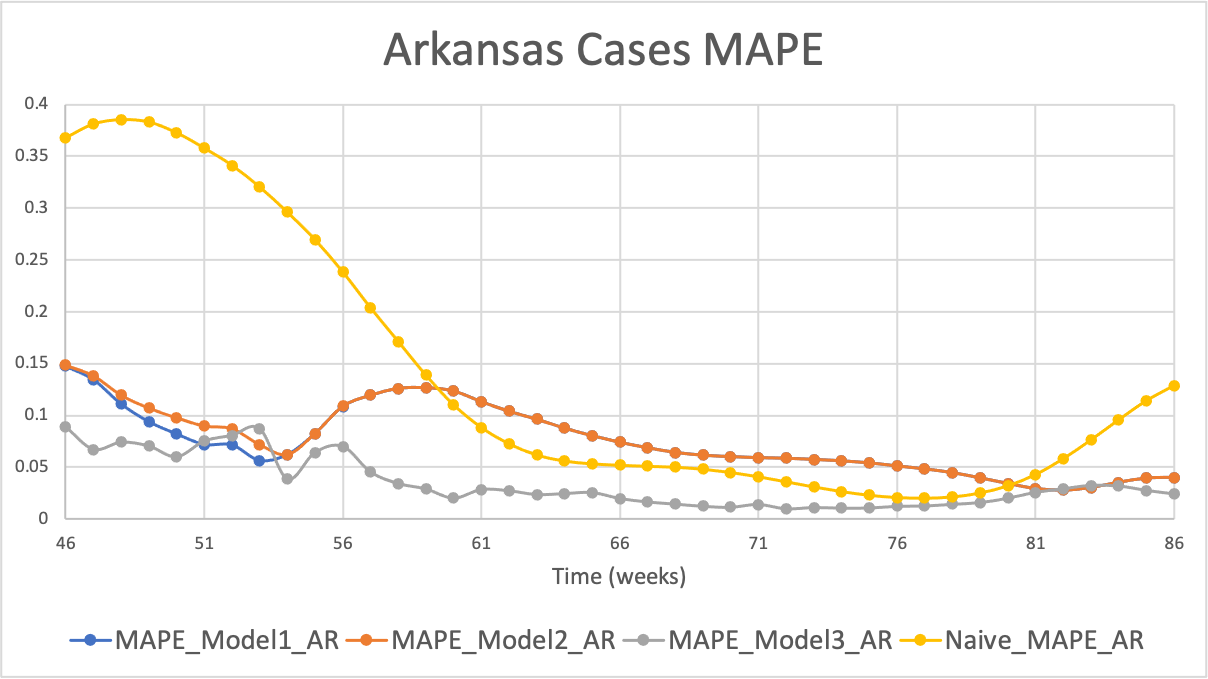}
\caption{Arkansas Cases {MAPE}. The vertical axis represents MAPE. Models 1 and 2 follow each other very closely, performing worse than the na\"ive model from approximately week 60 all the way to week 80. Models 1 and 2 experienced a steep increase in {MAPE} at week 55 and then trended downward gradually.}
\end{subfigure}
\begin{subfigure}[t]{0.45\textwidth}
\includegraphics[width=\textwidth]{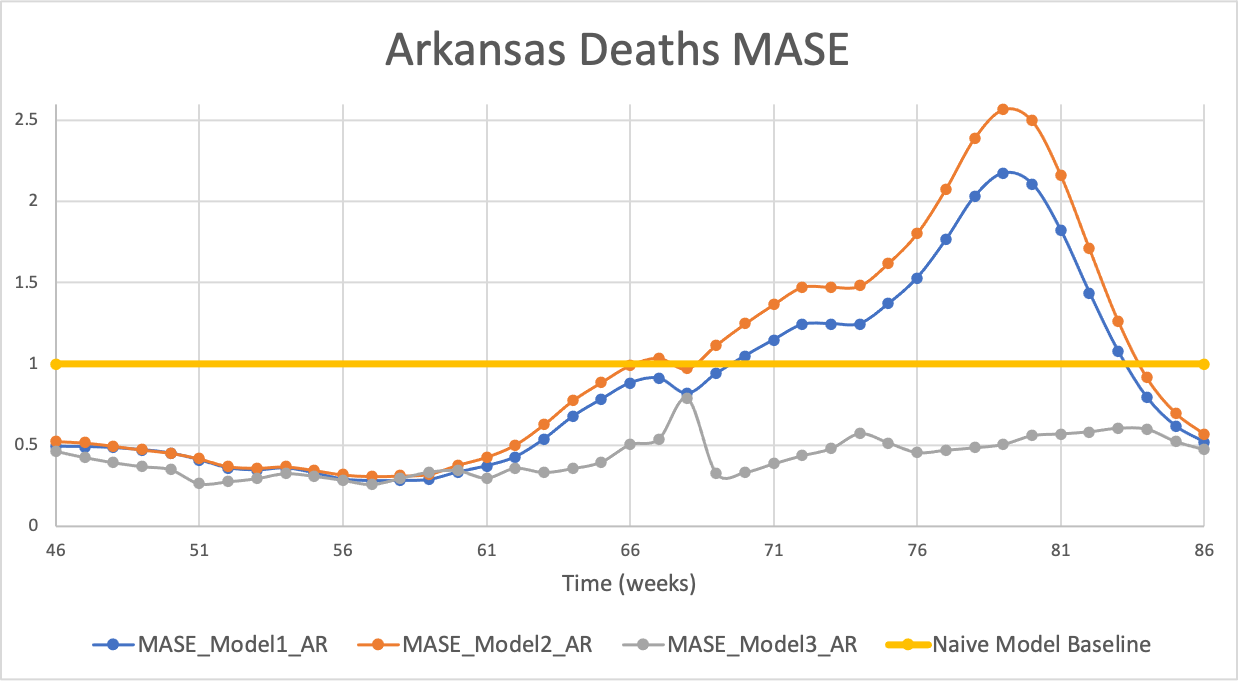}
\caption{Arkansas Deaths {MASE}. The vertical axis represents MASE. Models 1 and 2 follow each other closely. Models 1 and 2 start to perform worse than the na\"ive model at approximately week 69 but end up performing better than it at approximately week 83. This state does not experience a typical bimodal pattern of {MASE} as the other states.}
\end{subfigure}\hfill
\begin{subfigure}[t]{0.45\textwidth}
\includegraphics[width=\textwidth]{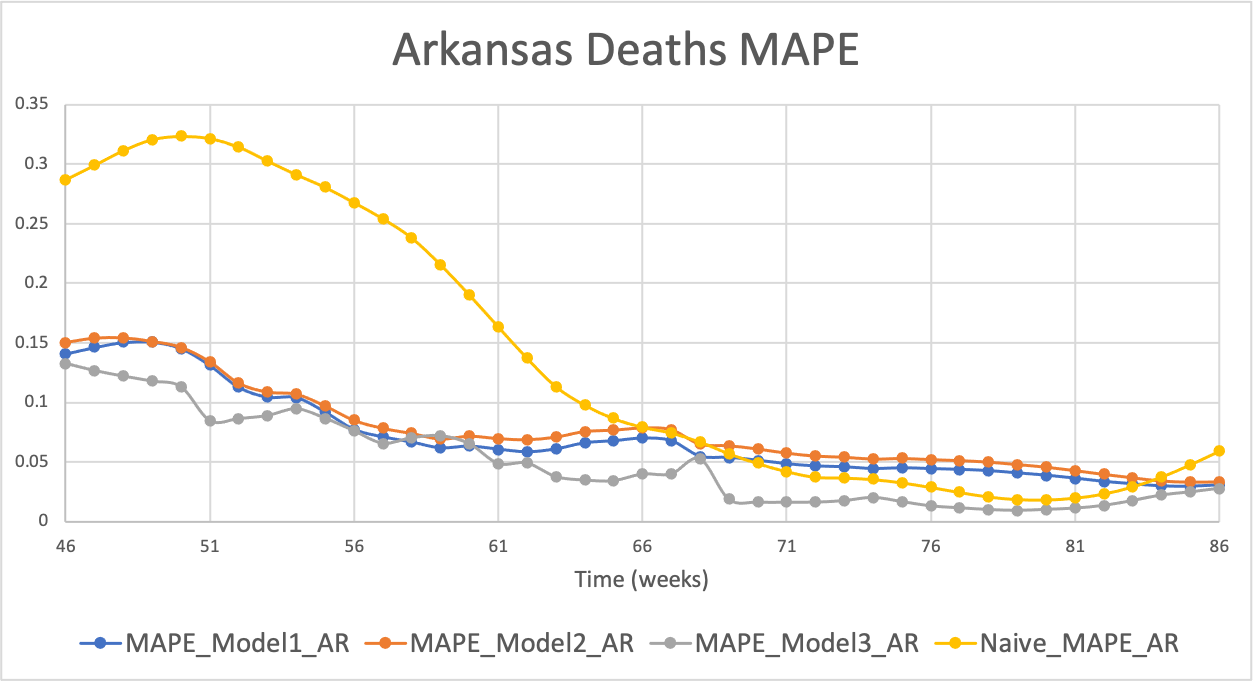}
\caption{Arkansas Deaths {MAPE}. The vertical axis represents MAPE. Models 1 and 2 follow each other very closely. They perform worse than the na\"ive model from approximately week 68 to week 83. All models trend gradually downwards.}
\end{subfigure}
\caption{The horizontal axis depicts the number of weeks since data collection began on January 22, 2020 for all plots in this figure. The four subfigures discuss the three models for the cases and deaths and the MAPE and MASE evaluation metrics for the state of Arkansas. Adapted from \cite{USCB:2019} and \cite{USAFacts:2021a}.
}
\label{fig:AR}
\end{figure}

Table \ref{table:summary} provides a summary of the results obtained for each of the models in each of the states. More details graphical results are shown in Figures \ref{fig:RI}--\ref{fig:AR} for Rhode Island (RI), Massachusetts (MA), California (CA), Florida (FL), and Arkansas (AR), respectively. {\color{black}Recall that in order for a model to have predictive power, its {MASE} must exceed the accuracy of the respective na\"ive model \cite{Lewis:1982}.  Therefore, in order for a model to have a predictive power, MASE should be smaller than 1. Additionally, a model's {MAPE} must be lower than 50\%. Table \ref{table:mape} describes appropriate interpretations for different levels of {MAPE}. 
\begin{table}[h]
{\color{black}
\caption[MAPE Interpretation]{\color{black}MAPE Interpretation. The following table provides a guide for interpretation of a time series model's {MAPE}. The higher the {MAPE}, the less predictive power the model has. Source: \cite{Lewis:1982}.}
\begin{center}
\begin{tabular}{|c|c|}
\hline
\textbf{MAPE}   & \textbf{Interpretation}     \\ \hline
\textless 10    & Highly accurate forecasting \\ \hline
10 - 20         & Good forecasting            \\ \hline
20 - 50         & Reasonable forecasting      \\ \hline
\textgreater 50 & Inaccurate forecasting      \\ \hline
\end{tabular}
\end{center}
\label{table:mape}
}
\end{table}
}

As can be seen from Table \ref{table:summary} and Fig.~\ref{fig:RI}, Rhode Island, a state with few counties but a dense population that is highly vaccinated, exhibits a great deal of difference among its models. Also from Table \ref{table:summary} and Fig.~\ref{fig:MA}, Massachusetts, similar to Rhode Island, has a more urban population that is highly vaccinated. This state did not exhibit much difference among its models; regardless, Model 3 outperformed all of them. California, a state with a large land area and population that is highly vaccinated, did not exhibit much difference among its models. Model 3 outperformed all of them, regardless as we can see from Table \ref{table:summary} and Fig.~\ref{fig:CA}. Florida, a state with a low vaccination rate and a large population, did not exhibit much difference among its models. Model 3 still performed the best out of all of them (Table \ref{table:summary} and Fig.~\ref{fig:FL}). Finally, Arkansas, a state with a great deal of rural land and one of the lowest vaccination rates in the country, performed differently than other states. As we see in Table \ref{table:summary} and Fig.~\ref{fig:AR}, Model 3 remains the best performing model.


\begin{table}[h]
\centering
\caption{Summary statistics of each model's performance in MAPE and MASE over the time horizon.}\label{table:summary}
\begin{tabular}{|l|llll|llll|} 
\hline
        & \multicolumn{8}{l|}{Rhode Island}                                                                                                                                                                    \\ 
\cline{2-9}
        & \multicolumn{4}{l|}{Cases}                                                                       & \multicolumn{4}{l|}{Deaths}                                                                       \\ 
\cline{2-9}
        & \multicolumn{2}{l|}{MASE}                                 & \multicolumn{2}{l|}{MAPE}            & \multicolumn{2}{l|}{MASE}                                 & \multicolumn{2}{l|}{MAPE}             \\ 
\cline{2-9}
        & \multicolumn{1}{l|}{Mean} & \multicolumn{1}{l|}{Variance} & \multicolumn{1}{l|}{Mean} & Variance & \multicolumn{1}{l|}{Mean} & \multicolumn{1}{l|}{Variance} & \multicolumn{1}{l|}{Mean} & Variance  \\ 
\hline
Model 1 & 0.8585                    & 0.7696                        & 0.0695                    & 0.0013   & 0.9888                    & 0.6436                        & 0.0615                    & 0.0009    \\
Model 2 & 1.129                     & 1.1376                        & 0.0906                    & 0.0019   & 1.8501                    & 2.0184                        & 0.1073                    & 0.0017    \\
Model 3 & 0.285                     & 0.0353                        & 0.0334                    & 0.0007   & 0.517                     & 0.0934                        & 0.0419                    & 0.0008    \\
Na\"ive   & 1                         & 0                             & 0.1642                    & 0.0205   & 1                         & 0                             & 0.1201                    & 0.0101    \\ 
\hline
        & \multicolumn{8}{l|}{Massachusetts}                                                                                                                                                                   \\ 
\cline{2-9}
        & \multicolumn{4}{l|}{Cases}                                                                       & \multicolumn{4}{l|}{Deaths}                                                                       \\ 
\cline{2-9}
        & \multicolumn{2}{l|}{MASE}                                 & \multicolumn{2}{l|}{MAPE}            & \multicolumn{2}{l|}{MASE}                                 & \multicolumn{2}{l|}{MAPE}             \\ 
\cline{2-9}
        & \multicolumn{1}{l|}{Mean} & \multicolumn{1}{l|}{Variance} & \multicolumn{1}{l|}{Mean} & Variance & \multicolumn{1}{l|}{Mean} & \multicolumn{1}{l|}{Variance} & \multicolumn{1}{l|}{Mean} & Variance  \\ 
\hline
Model 1 & 0.8579                    & 0.5338                        & 0.0723                    & 0.0013   & 0.8263                    & 0.2572                        & 0.0691                    & 0.0015    \\
Model 2 & 0.9869                    & 0.7752                        & 0.078                     & 0.0012   & 0.8275                    & 0.2586                        & 0.0692                    & 0.0015    \\
Model 3 & 0.2556                    & 0.0193                        & 0.0327                    & 0.0008   & 0.3933                    & 0.0103                        & 0.0467                    & 0.0014    \\
Na\"ive   & 1                         & 0                             & 0.1582                    & 0.0204   & 1                         & 0                             & 0.1374                    & 0.0136    \\ 
\hline
        & \multicolumn{8}{l|}{California}                                                                                                                                                                      \\ 
\cline{2-9}
        & \multicolumn{4}{l|}{Cases}                                                                       & \multicolumn{4}{l|}{Deaths}                                                                       \\ 
\cline{2-9}
        & \multicolumn{2}{l|}{MASE}                                 & \multicolumn{2}{l|}{MAPE}            & \multicolumn{2}{l|}{MASE}                                 & \multicolumn{2}{l|}{MAPE}             \\ 
\cline{2-9}
        & \multicolumn{1}{l|}{Mean} & \multicolumn{1}{l|}{Variance} & \multicolumn{1}{l|}{Mean} & Variance & \multicolumn{1}{l|}{Mean} & \multicolumn{1}{l|}{Variance} & \multicolumn{1}{l|}{Mean} & Variance  \\ 
\hline
Model 1 & 0.9056                    & 0.3836                        & 0.0764                    & 0.0014   & 0.801                     & 0.2                           & 0.0743                    & 0.0019    \\
Model 2 & 0.984                     & 0.5088                        & 0.083                     & 0.0015   & 0.9192                    & 0.2475                        & 0.084                     & 0.0019    \\
Model 3 & 0.4199                    & 0.0403                        & 0.0492                    & 0.0023   & 0.47                      & 0.0169                        & 0.0594                    & 0.0028    \\
Na\"ive   & 1                         & 0                             & 0.1507                    & 0.0197   & 1                         & 0                             & 0.1447                    & 0.0152    \\ 
\hline
        & \multicolumn{8}{l|}{Florida}                                                                                                                                                                         \\ 
\cline{2-9}
        & \multicolumn{4}{l|}{Cases}                                                                       & \multicolumn{4}{l|}{Deaths}                                                                       \\ 
\cline{2-9}
        & \multicolumn{2}{l|}{MASE}                                 & \multicolumn{2}{l|}{MAPE}            & \multicolumn{2}{l|}{MASE}                                 & \multicolumn{2}{l|}{MAPE}             \\ 
\cline{2-9}
        & \multicolumn{1}{l|}{Mean} & \multicolumn{1}{l|}{Variance} & \multicolumn{1}{l|}{Mean} & Variance & \multicolumn{1}{l|}{Mean} & \multicolumn{1}{l|}{Variance} & \multicolumn{1}{l|}{Mean} & Variance  \\ 
\hline
Model 1 & 0.9679                    & 0.5556                        & 0.0735                    & 0.0011   & 0.8444                    & 0.3151                        & 0.0693                    & 0.0015    \\
Model 2 & 1.0033                    & 0.598                         & 0.0754                    & 0.001    & 0.9405                    & 0.4205                        & 0.0743                    & 0.0014    \\
Model 3 & 0.3098                    & 0.0272                        & 0.0365                    & 0.0012   & 0.4182                    & 0.0119                        & 0.0491                    & 0.0014    \\
Na\"ive   & 1                         & 0                             & 0.145                     & 0.0177   & 1                         & 0                             & 0.1382                    & 0.0132    \\ 
\hline
        & \multicolumn{8}{l|}{Arkansas}                                                                                                                                                                        \\ 
\cline{2-9}
        & \multicolumn{4}{l|}{Cases}                                                                       & \multicolumn{4}{l|}{Deaths}                                                                       \\ 
\cline{2-9}
        & \multicolumn{2}{l|}{MASE}                                 & \multicolumn{2}{l|}{MAPE}            & \multicolumn{2}{l|}{MASE}                                 & \multicolumn{2}{l|}{MAPE}             \\ 
\cline{2-9}
        & \multicolumn{1}{l|}{Mean} & \multicolumn{1}{l|}{Variance} & \multicolumn{1}{l|}{Mean} & Variance & \multicolumn{1}{l|}{Mean} & \multicolumn{1}{l|}{Variance} & \multicolumn{1}{l|}{Mean} & Variance  \\ 
\hline
Model 1 & 0.9679                    & 0.5556                        & 0.0735                    & 0.0011   & 0.8573                    & 0.3106                        & 0.0698                    & 0.0014    \\
Model 2 & 1.0033                    & 0.598                         & 0.0754                    & 0.001    & 0.9903                    & 0.4565                        & 0.0767                    & 0.0013    \\
Model 3 & 0.3098                    & 0.0272                        & 0.0365                    & 0.0012   & 0.424                     & 0.0141                        & 0.0496                    & 0.0014    \\
Na\"ive   & 1                         & 0                             & 0.145                     & 0.0177   & 1                         & 0                             & 0.138                     & 0.0136    \\
\hline
\end{tabular}
\end{table}

All states analyzed exhibit similar behavior in the measures of performance via the {MAPE} and {MASE}. All models decrease gradually in the {MAPE} as the time series prediction passes through the time horizon. The {MAPE} for the na\"ive model consistently starts very high but then gradually approaches 0 along with the other models. Models 1 and 2 perform worse than the {MAPE} for the na\"ive model from approximately week 71 to week 84. Most of the states exhibited a bimodal structure in the {MASE} from approximately week 66 to week 84. Approximately week 70 to week 77 showed a rapid increase in the {MASE} across all states. Models for all states perform similarly because of the inclusion of both into-state and out-of-state travel. Likely, there are some counties that people throughout the country commute to that are included in many states, including those that we select in this work. Because Model 3 outperforms the na\"ive model most of the time in the time horizon for all states, this model should be considered by epidemiologists. The {GNAR} models on county networks with commuting information prove potentially useful for predicting {COVID-19} cases and deaths.

\section{Conclusion}
\label{sec:conclusion}

The coronavirus pandemic has ravaged the world, killing many, and affecting the daily lives of all people. Concentrated efforts from all parts of the earth have attempted to curb this virus' spread. From mathematical models to public health policy decisions, these efforts have brought the world together in an attempt to eradicate this virus. 
In this work, we show that the {GNAR} model performs very well in predicting {COVID-19} cases and deaths throughout the county network in the United States. Using the open-source data from common sources, including the {USCB} and {USAFacts}, we can create a predictive model that could better inform public health officials. 

For example, {\color{black}c}ell phone data is both nearly ubiquitous and surprisingly accurate \cite{Johnson:2021}. Companies and organizations have been able to harness the data from commuter's cell phones using their navigation applications in order to better influence their prediction of traffic flow through an area \cite{Johnson:2021}. This data is almost live, since it comes directly from the drivers as they drive along a road.
This live traffic data could help describe a by-county commuting network. The network could be dynamic, changing as more data is obtained. Perhaps analyzing trends over the past few weeks in a local area could result in a more accurate and current county commuting network structure.
{\color{black}In addition, in this paper, we assume that the traffic and commuting patterns by county remain the same through the time of the {COVID-19} pandemic as of the {USCB} compiled this commuting data over a five year period from 2011 to 2015, giving it a static property. In the beginning of the pandemic, because of the lock-down in many states, we had much less traffic flows.  The effect of less traffic flows between counties might be most to Model 1 and Model 2 since Model 1 and Model 2 put parameters to weight on neighbor traffic flows.  However, since Model 3 has less weight on the traffic flows between a county and its neighbor counties,  Model 3 was less affected by the change of traffic flows between counties.  If we had information on traffic flows between counties during the pandemic, we expect that the performances of Model 1 and Model 2 will be improved.  }

The data sources of this work are delineated by county, which provides data for a more localized area. However, one could further subdivide this data into zone improvement plan (ZIP) codes in order to obtain an even further refined prediction at a lower level. The {CDC} currently only collects data at the county level; however, with future technologies for tracking a disease's spread, the {CDC} could subdivide its data even further. As of November 2021, there exist 41,692 {ZIP} codes in the United States \cite{USPS:2021}. Since individuals are freely able to move between their {ZIP} codes, and since the frequency of moving between {ZIP} codes is likely higher on average, this subdivision of data may provide a great deal of insight into localized trends of movement of people. {ZIP} code analysis may demonstrate a more realistic representation of daily life and community interactions due to the relatively smaller distance between nodes.

{\color{black}In this paper we treated all states the same so that we ignored covariates $c = 1, \ldots , C$ in Equation (\ref{eq:gnar}).  We might be able to use covariates $c = 1, \ldots , C$ for information of low and high vaccination rates in states and it might increase performances of the models using the GNAR model to predict the number of cases.  }

Finally, one could find any network structure and incorporate it into the {GNAR} model as long as it is geographically delineated the same way as the time series data. Any data that describes a flow from one geographic area to another can be formulated into a network structure, which is a key component of a {GNAR} model. Comparing multiple network structures could provide insight into what is important in the dissemination of a disease. With the advent of structure centrality \cite{rasti2021novel}, this could be a possibly interesting avenue for extending traditional centrality metrics (see, e.g., \cite{sarlas2020betweenness}) in epidemic spreading.

Applying this methodology to other geographic areas or governance divisions could also prove useful around the world, not just the United States. Any country's municipalities, provinces, or townships could represent nodes in a network similar to the United States county structure. Comparing countries of a similar geographic, climatic, and demographic makeup to the United States may especially prove insightful. One can also compare and contrast the public health policy effects in different geographic areas.

\section*{Reproducibility and code availability}

A dashboard with all of the results presented here is available through \url{http://chvogiat.shinyapps.io/dod2021}: this version was also presented during the Dynamics of Disasters 2021 conference in June 2021. All codes and data are available upon request.

\section*{Data availability}

The datasets analyzed during the current study are available from the corresponding author on reasonable request. All data are also publicly available from the United States Census Bureau \cite{USCB:2019}, the United States of America Facts \cite{USAFacts:2021a}, and the Center for Disease Control and Prevention \cite{CDC:2020}.

\section*{Author contributions}
PU and DW are both considered as first authors of this work and hence are listed in alphabetical order; CV is the corresponding author; RY is the supervisor of this research. All authors equally contributed to this research.  
%
\section*{Conflict of interest}
The authors declare that they have no conflict of interest.

\section*{Acknowledgments}
We would like to thank the two anonymous reviewers, as well as the editor, for their comments that helped us improve the work from its first iteration. We would also like to take this opportunity to thank and acknowledge all workers who provided essential services during this COVID-19 pandemic.

\bibliographystyle{spmpsci}      
\bibliography{citations}   

%
%

\end{document}